\documentclass[10pt,journal,final,onecolumn]{IEEEtran}
\usepackage[utf8]{inputenc}
\usepackage{amsmath}
\usepackage{amsfonts}
\usepackage{amssymb}
\usepackage{graphicx}
\usepackage{float}
\allowdisplaybreaks
\usepackage{multirow,array}
\usepackage{mathrsfs}
\usepackage{float}
\usepackage{comment}

\usepackage{amsmath}
\usepackage{amsfonts}
\usepackage{amssymb}
\usepackage{graphicx}
\usepackage{float}
\usepackage{longtable}
\usepackage{multirow,array}
\usepackage{float}

\usepackage{fancyhdr}
\usepackage{bm}
\usepackage{amsmath}

\usepackage{booktabs}
\usepackage{adjustbox}
\usepackage{longtable}
\usepackage{longtable,lscape}
\usepackage{rotating}
\usepackage[graphicx]{realboxes}
 
\usepackage{xcolor}

\usepackage{fancyhdr}
\usepackage{amsmath}
\usepackage{amsfonts}
\usepackage{amssymb}
\usepackage{graphicx}
\usepackage{float}

\usepackage{multirow,array}
\usepackage{float}
\usepackage{fancyhdr}
\usepackage{bm}
\usepackage{amsmath}

\usepackage{booktabs}
\usepackage{adjustbox}
\usepackage{longtable}
\usepackage{longtable,lscape}
\usepackage{rotating}
\usepackage[graphicx]{realboxes}
\usepackage{amsmath}

\usepackage{booktabs}
\usepackage{adjustbox}
\usepackage{longtable}
\usepackage{longtable,lscape}
\usepackage{rotating}
\usepackage[graphicx]{realboxes}
\begin{document}
	
	
	%
	\setlength\parindent{2em}

		\title{A general joint latent class model of longitudinal and survival data with the covariance modelling }
	\author{Ruoyu Miao; Christiana Charalambous  \\
		The University of Manchester
	}
	\date{2020/12/13}  
	\maketitle  
	\begin{abstract}
		\quad Based on the proposed time-varying JLCM (Miao and Charalambous, 2022), the heterogeneous random covariance matrix can also be considered, and a regression submodel for the variance-covariance matrix of the multivariate latent random effects can be added to the joint latent class model.  A general joint latent class model with heterogeneous random-effects modelling is a natural extension of the time-varying JLCM, which consists of the linear and the log link functions to model the covariance matrices as the variance-covariance regression submodel based on the modified Cholesky decomposition, longitudinal submodel, survival submodel as well as the membership probability. It can help to get more information from the random covariance matrix through the regression submodel and get unbiased estimates for all parameters by modelling the variance-covariance matrix. By adding the regression model, the homogeneous random effects assumption can be tested and the issue of high-dimensional heterogeneous random effects can be easily solved. The Bayesian approach will be used to estimate the data. DIC value is the criterion for deciding the optimal $k$ value. We illustrate our general  JLCM on a real data set of AIDS study and we are interested in the prospective accuracy of our proposed JLCM as well as doing the dynamic predictions for time-to-death in the joint model using the longitudinal CD4 cell count measurements.\\
		
	\end{abstract}
	
	\section{Introduction}
	\indent \quad In the statistical literature, the random-effects model is usually used to dealwith the repeated measures (Laird and Ware, 1982) and the  Cox proportional hazards models are widely used to analyze the survival data and  estimate the treatment effect on the time-dependent covariate (Cox, 1972).  However, separate analyses  of longitudinal and time-to-event outcomes can  lead to  biased estimation  when the  time-to-event process contains  endogenous time-varying covariates, or both  outcome processes are correlated.  Ibrahim et al. (2010) pointed out that it will lead to the bias to treatment effects when the longitudinal measurements are ignored. They also proved that the joint model can help to provide more efficient estimates of the treatment effects on longitudinal and survival processes.  Joint models are proposed to improve the inference of survival analysis with time-varying covariates measured with error, and analyze the relationship between longitudinal measurements and time-to-event process. What is more, even in the case of just longitudinal data, there could  be issues with informative dropouts,  which can also be addressed by considering the time to dropout in the joint modelling context.  Under these situations, joint modelling of both longitudinal and time-to-event data can incorporate all information, increasing the efficiency and decreasing the bias of  inferences (Ibrahim et al., 2010).\\
\indent \quad The early work on joint modelling dates back to the period from the late 80s to the early 90s. Schluchter (1992) pointed out the problem of  non-ignorable censoring in  longitudinal studies and gave an approach which was based on  lognormal survival. In AIDS research, Self and Pawitan (1992), Degruttola and Tu (1994) and Tsiatis et al. (1995)   adjusted the inferences on the longitudinal responses with informative missing data and considered a  two-stage method to do the parameter estimations. The two-stage approach splits the estimation of joint modelling into two steps. First, the model is  fitted for the longitudinal outcomes, and then the estimation outputs from the first step are used to fit the survival model. Although the two-stage method is easy to implement, it has the limitation of leading to  biased parameter estimation because of the omission of the informatively censored events.\\
\indent \quad
Fawcett and Thomas (1996) gave the structure of a joint model as a covariate tracking model with measurement error for the longitudinal submodel and the proportional hazards model for the survival submodel. In this work, Gibbs sampling was used for  statistical inference. Moreover, Wulfsohn and Tsiatis (1997) introduced the standard joint model composed by the linear growth curve model with random intercept and slope and the proportional hazards model, using the expectation maximization (EM) algorithm to estimate the parameters. This standard structure of joint modelling with time-dependent relative risk also improved on two-stage methods.  Empirical Bayes estimates of the covariate process were computed in the first stage and also treated as the time-dependent covariates and  used  in the second stage to get  parameters  which maximize the partial likelihood for the Cox model.  Additionally, Henderson et al. (2000) proposed a flexible joint model, where the longitudinal and survival processes were linked via a latent Gaussian process, allowing for both serial correlation and measurement error in the longitudinal model.  Albert et al. (2000) presented the joint model for longitudinal binary data with the shared Gaussian autoregressive latent process.  \\
\indent \quad  Tsiatis and Davidian (2004) provided an overview of the early work of joint modelling over the last two decades and described two approaches: the conditional score method and the semiparametric likelihood method. Wang (2006) proposed the corrected score method and did the comparison with the conditional score method, naive estimator as well as  regression calibration (RC). They found that  under general situations, the  RC method can significantly reduce biases from the naive estimator and the conditional score method outperforms the corrected score method slightly but the difference  of these two methods is small.  Rizopoulos (2009) proposed a new computational approach, in the form of a fully exponential Laplace approximation for the joint modelling of survival and longitudinal data, making is feasible to handle the high dimensional random effects structures in joint models. Motivated by primary biliary cirrhosis (PBC) study (Murtaugh et al., 1994), a joint modelling of multi-state event times and longitudinal data with informative observation time points was proposed by Dai and Pan (2018), in  which the joint modelling was linked by random effects under no distribution assumption. This model helps to extend the corrected score method to cases with longitudinal data which is collected at informative time points.\\
 \indent \quad Tseng et al. (2005) explored the joint modelling of the linear mixed effects model for longitudinal outcomes and the accelerated failure time model for the survival data under the EM algorithm to estimate the unknown parameters. They indicated that the accelerated failure time model could be considered an alternative to the Cox proportional hazards model when the assumption of proportionality failed to describe the relationship between the longitudinal and survival covariates.  Viviani et al. (2014) proposed the generalized linear mixed joint models (GLMJMs) under the estimation approach of the EM algorithm. This kind of joint modelling provided methods to control the  non-Gaussian longitudinal responses in the joint modelling frameworks. A Bayesian approach was also considered to robust joint modelling of longitudinal measurements and time-to-event  outcomes  by  Baghfalaki et al. (2014). In this approach, Students' $t$ distribution was applied to deal with outlier points. Among these various types of joint models, the typical and widely used submodel settings are the mixed-effects model for the longitudinal measurements and the Cox proportional hazards model for the time-to-event data, linked together by the shared random effects.  Further extensions to consider multiple longitudinal outcomes and/or competing risks (Gichangi \& Vach, 2005; Elashoff et al., 2008; Hu et al., 2009; Huang et al., 2011; Proust-Lima et al., 2015; Rouanet et al., 2016) have also been considered.  Other work in the literature includes copula joint models (K{\"u}r{\"u}m et al., 2018, Zhang et al., 2021) as well as hidden Markov models (Bartolucci \& Farcomeni, 2019; Zhou et al., 2022).\\
  \indent \quad  Hickey et al. (2016)  gave a review of joint models with frequentist approaches, with the expectation maximization method, Bayesian approach for estimation    and other estimation approaches  also illustrated briefly.    More recently, Alsefri et al. (2020) presented a methodological review for  Bayesian univariate and multivariate joint modelling of longitudinal and survival analysis, including model structures, estimation procedures, dynamic predictions as well as software implementation.\\
    \indent \quad  The development of specialised software, such as JM (Rizopoulos, 2010), joineR (Philipson et al., 2012), stjm (Crowther et al., 2013) and JMBayes (Rizopoulos, 2016) among others, has also helped popularise the use of joint models in practice. The JM package (Rizopoulos, 2012) and joineR (Philipson et al., 2012) were developed using the likelihood maximization estimation method for joint model. Rizopoulos  (2016) proposed shared parameter models for the joint modelling of both longitudinal and time-to-event data using MCMC procedure with the JMbayes package.\\
        \indent \quad The basic concept of latent class (LC) modelling  was originally   introduced by  Lazarsfeld (1950)   for creating typologies (or clusters) from dichotomous variables as part of his more comprehensive latent structure analysis.   LC modelling was extended by Goodman (1974a, 1974b) with MLE methods, which the previous implementation problems were resolved. Haberman (1979) demonstrated the relationship between LC models and log-linear models. Hagenaars (1990) and Vermunt (1993) developed a general  framework for categorical data analysis with discrete latent variables (1997). LC models have been applied widely (Fergusson\&  Horwood, 1989; Fergusson et al., 1991; 
        Fergusson et al., 1994; Bucholz, 1996)  and have been developed for longitudinal data with a Markov model (van de Pol \& Langeheine, 1989).\\
            \indent \quad  Reboussin and Anthony (2001) proposed the  dynamic  latent class regression model for longitudinal data, which could allow the conditional probabilities to vary over time.  They added the   time-varying covariates into the baseline-category logistic regression model, which was the first instance of time-varying probability model for the membership probability. Based on the  time-varying latent class regression model proposed by Reboussin and Anthony (2001),  Lin et al. (2014) added random effects to the  dynamic latent class model for longitudinal data, which they  jointly modelled  the informative event with a class-specific logistic model with shared random effects. \\
                \indent \quad  One disadvantage of the basic joint models is that they cannot handle data with potential heterogeneous  subgroups.  Due to the issue of the existence on underlying patterns of subjects in many clinical trials (e.g. renal transplantation data with two underlying subpopulation for RC levels (Garre et al., 2008) and CPCRA AIDS  study (Abrans et al., 1994; Neaton et al., 1994) with three latent classes for CD4 counts (Liu et al., 2015)),   the joint latent class model (JLCM) is  highly regarded as an important contribution to the joint modelling literature,    firstly proposed by Lin et al. (2002) to deal with the  non-ignorable missing data in longitudinal measurements and  identify the pattern of longitudinal data over time.  A multinomial logistic model with class-specific coefficients was considered as the membership submodel to do the subgroup specification, which was added to the joint modelling of longitudinal biomarkers and the event process. These three submodels were linked by the latent class number and the EM algorithm was used to do the estimation. \\
                \indent \quad       Joint latent class modelling of longitudinal measurements and time-to-event data  with Bayesian  approach  has also been developed recently (Entink et al., 2011;  Chen and Huang, 2015; Huang et al., 2016; Dagne, 2017; Andrinopoulou  et al., 2020). Huang et al. (2016) proposed a mixture of nonlinear mixed-effects joint (MNLMEJ) models for longitudinal  and survival analysis with  Bayesian approach. This MNLMEJ model presented three different  processes (response, covariate as well as time-to-event), which three submodels were linked by the shared random effects and the membership probability was obtained from Dirichlet distribution. Andrinopoulou  et al. (2020) applied the cystic fibrosis dataset into the joint model of longitudinal and survival outcomes with latent classes which were sampled from Dirichlet distribution, combining   a new optimal latent class selection procedure with a mixture model assuming many more latent classes than present in the data. Unlike the classic frequentist estimation approach,  they  presented a Bayesian approach for the shared parameter joint latent class model.  The overfitted mixture model was considered to identify the non-empty and  the empty classes to select the optimal number of classes. Different from the membership probability based on the logistic model, they assumed each individual has a probability of being to the latent class which is obtained using a multinomial distribution.\\
                     \indent \quad  Zhang and Simonoff (2020)  discussed both the strengths and weaknesses of the joint latent class modelling (JLCM) and presented an idea of  nonparametric joint modelling approach. They pointed that  the joint latent class tree (JLCT) model  with a tree-based approach could be a good alternative to JLCM, which can addresses the time-invariant limitation of JLCM.  Zhang and Simonoff (2022) proposed a tree-based approach to model time-to-event and longitudinal data. This semiparametric joint latent class tree model is fast to fit and can allow time-varying covariates for all components in modelling time-to-event and latent class membership, which the time-varying covariates can be used as the splitting variables to construct tree. However, this JLCT   cannot show how memberships of subjects change with time in details.\\              
                      \indent \quad   Motivated by Garre et al. (2008) and Liu et al. (2015), the subpopulations  cannot be ignored in subjects. For example, treatment A, which might be effective for patient a could prove ineffective  for patient b, which indicates classification can make a huge contribution to the data analysis depending on the characteristics of the dataset. 
                     Under this situation, patients  may experience pathological changes such that a particular treatment regime might stop being effective for a patient and a different regime might need to be considered. This could be reflected by a change in the latent class for that patient. In order to handle this kind of particular dataset with jumping behaviours, we propose adding time-varying covariates into the latent class membership probability, which allows jumps among different subgroups (Miao and Charalambous, 2022). \\    
                      \indent \quad  Covariance modelling also has a long history of development over the past few years.  For many years, the variances/covariances were considered as the ‘nuisance parameter’  with respect to the mean when modelling  the repeated measures on the same individual (MacKenzie, 2004). For the classical  ANOVA and multivariate approaches, they both ignore the covariance structure, which can have a bad impact on the efficiency of the mean parameters. Today, however,  variance-covariance modelling   is equally as important as the mean  modelling.
                     Anderson (1973)  proposed a covariance model with  linear structure, in which positive definiteness could not be guaranteed.  Leonard and Hsu (1992) proposed a Bayesian approach to model the covariance matrix in a multivariate Normal distribution. Pinheiro and Bates (1996) and Chiu et al. (1996) considered various covariance parameterisations, including the Cholesky decomposition, which could guarantee the positive definiteness of the covariance matrix, leading to unconstrained estimation.     Pourahmadi (1999) firstly proposed  the modified Cholesky decomposition to get the positive-definiteness of the covariance matrix with unconstrained parameterizations, which was generalized to fit unbalanced longitudinal data by Pan and MacKenzie (2003).  Pourahmadi (1999, 2000) proposed a modified Cholesky decomposition to reparameterize the inverse of the covariance matrix with $T^{T}D^{-1}T$, where $T$ denotes  a unique unit lower triangular matrx with  1's 
                     as diagonal entries and $D$ is  a unique diagonal matrix  with positive diagonal entries. Based on the modified Cholesky decomposition proposed by  Pourahmadi (1999, 2000), Pan and MacKenzie (2006) modelled a marginal covariance, while modelling the conditional covariance for the linear mixed model was considered by
                     Pan and MacKenzie (2007).\\
                      \indent \quad  According to the modified Cholesky decomposition, Huang et al. (2010) proposed a joint model of longitudinal and competing risks survival data with heterogeneous random effects and outlying longitudinal measures, which indicated that $t$-distribution can help to deal with the potential  outliers in the longitudinal measurements.  And this was the first instance of covariance modelling being applied in joint models. They indicated that it would lead to bias parameter estimates  for the survival endpoint if covariance modelling was ignored and unbiasd estimated results can be obtained by the  heterogeneous covariance matrix of the multivariate random effects with covaraince modelling. \\                   
\indent \quad  No covariance modelling has been proposed in the context of joint latent class models in previous work. Motivated by the literature in covariance modelling,  we 
add covariance modelling to the time-varying JLCM. Based on the time-varying joint latent class model (JLCM) we mentioned before,   we  propose  a general JLCM of longitudinal measurements and time-to-event outcomes with both time-varying membership probability as well as the covariance modelling, which the regression submodel is added to get more information in the variance covariance matrix.    However, we do not have the same structure with the  shared parameter in our proposed time-varying JLCM.  We use the shared multivariate distribution for random effects in the longitudinal and survival submodels for the general JLCM.  A Bayesian approach is used for estimation and inference. The detailed structure of the general JLCM  is presented in section \textrm{II}. In section \textrm{III}, we conduct a simulation study to compare the performance of the proposed JLCM with the time-varying JLCM, while in section  \textrm{IV}, we apply our model to analyse the AIDS dataset (Goldman et al., 1996).   For both simulation and real data analyses, we also investigate the performance of the proposed model in the dynamic prediction of survival probabilities. Finally, section  \textrm{V} summarises the work in this paper and discusses future extensions of the proposed model. \\
	\section{Joint latent class model with time-varying probability and heterogeneous random effects}	
	\quad   Based on the improved time-varying JLCM with shared random effects proposed (Miao and Charalambous, 2022), a general joint latent class model with heterogeneous random effects modelling, is considered. This model is motivated by Huang et al. (2010) who proposed a joint model of longitudinal and competing risks survival data with heterogeneous random effects, with additional modelling of the covariance matrix of the random effects. 
	 As opposed to the time-varying JLCM with  the shared random effects we proposed before, we assume that  the random effects in the general JLCM are different for each process but share the same multivariate distribution. \\
	\indent\quad In this section, we proposed a general  joint latent class model  of longitudinal and survival data with  the time-vaying membership probability and covariance modelling.   Compared with basic joint latent class model,  we add time-varing covariates into latent class submodel to get time-varying probability which we have proposed before. Here as the comparable JLCM, we develop the similar time-varying JLCM but the structure is different, which the shared distribution is used to link longitudinal and survival submodels. 	The detailed structure of the general JLCM and  estimation method will be illustrated. A simulation study  will be conducted to compare the performance of the proposed general JLCM with a time-varying JLCM. The proposed general JLCM is used to analyse the AIDS  dataset (Goldman et al., 1996).   For both simulation and real data analyses, we also investigate the performance of the proposed model in the dynamic prediction of survival probabilities.\\
	\subsection{General latent class model with time varying probability}
	\quad Here we  model a general submodel which varies with time $t$. Since  the value of  the membership probability  is between 0 and 1, we consider  the logistic model.  Consider a scenario with $N$ subjects, indexed by $i = 1,2, \dots, N$, and $K$ latent classes, indexed by $k = 1,2, \dots, K$. The membership probability is assumed to vary over time at measurement points $t$. Let the design matrices $\bm{X}_{1i}$ incorporate time-dependent covariates observed at specific time points $t_i = t_{ij} \big|_{j=1,2,\dots,m_i} = (t_{i1}, t_{i2}, \dots, t_{im_i})^T$.  Define $R_{ij} = k$ as the latent class indicator, signifying that subject $i$ belongs to class $k$ at time $t_{ij}$. The probability $\pi_{ijk}$ that subject $i$, measured at time $t_{ij}$, belongs to class $k$ satisfies the constraint $\sum_{k=1}^{K} \pi_{ijk} = 1$ for all $k = 1,2,\dots,K$. This probability is modeled using a multinomial logistic regression, incorporating the covariate vector $\textbf{X}_{1i}$ along with the corresponding class-specific coefficient vector $\bm{\xi}_k$.  Thus, we present a generalized latent class framework where the probability of subject $i$ belonging to class $k$ at their $j$th visit is influenced by time-dependent covariates, as given by:
\\ 
	$$\begin{aligned}
		\pi_{ijk}=P(R_{ij}=k)=\frac{\exp(\bm{X}_{1i}^T(t_{ij})\bm{\xi}_{k})}{\sum\limits_{s=1}^{K}\exp(\bm{X}_{1i}^T(t_{ij})\bm{\xi}_{s})} \  \  \ \forall~~k=1, \ldots, K
	\end{aligned}  
	\eqno{(1)}
	$$

	\subsection{Longitudinal submodel}
	\indent\quad  Let $y_{ijk}$ represent the repeated measurement for subject $i$ at the $j$th visit within class $k$. The longitudinal submodel describing $y_{ij}$ follows the framework proposed by Troxel et al. (1998):
	$$\begin{aligned}
		y_{ij}|(R_{ij}=k)=\bm{X}_{2i}^T(t_{ij})\bm{\beta}_{k}+\bm{Z}_{i}^T(t_{ij})\bm{U}_{ik}+\epsilon_{ijk}
		\ \ \ \ \ K \ne 1
	\end{aligned}  
	$$
where $\bm{X}_{2i}(t_{ij})$ denote the covariate vector for subject $i$ at the $j$th visit time $t_{ij}$ and time is incorporated as a polynomial term in $\bm{X}_{2i}$. The parameter vector $\bm{\beta}_{k}$ corresponds to class $k$ at each visit time. The class-specific random effects for class $k$ are represented by $\bm{U}_{ik}$, which is a $q \times 1$ vector. We define $\bm{Z}_{i}(t_{ij})$ as the covariate vector at time $t_{ij}$ that is associated with the class-specific random effects $\bm{U}_{ik}$. Both $\bm{X}_{2i}(t_{ij})$ and $\bm{Z}_{i}(t_{ij})$ may contain time-dependent covariates measured at multiple time points, given by $t_i= (t_{i1}, t_{i2}, \dots, t_{im_i})^T$.  The class-specific error term $\epsilon_{ijk}$ is assumed to follow an independent normal distribution, $\epsilon_{ijk} \stackrel{\text{i.i.d.}}{\sim} N(0,\tau_k)$, with variance $\tau_k$. Furthermore, we assume that $\epsilon_{ijk}$ is independent of $\bm{U}_{ik}$ (Liu et al., 2015).  The random effects $\bm{U}_{ik}$ can be alternatively expressed as $\bm{U}_{ik}=\mathfrak{Z}_k\bm{U}_i$, where $\bm{U}_{ik} \sim N_{q}(0, \Sigma_{u_{ik}})$. For simplification, we assume $\mathfrak{Z}_k=1$, leading to $\bm{U}_{i} \sim N_{q}(0, \Sigma_{u_{i}})$, which implies that the random effects remain consistent across different latent classes. Under this assumption, the longitudinal submodel can be rewritten as:
	$$\begin{aligned}
		y_{ij}|(R_{ij}=k)=\bm{X}_{2i}^T(t_{ij})\bm{\beta}_{k}+\bm{Z}_{i}^T(t_{ij})\bm{U}_{i}+\epsilon_{ijk}
		\ \ \ \ \ K \ne 1
	\end{aligned}  
	\eqno{(2)}
	$$\\
	\subsection{Cause-specific hazard distribution}
	\indent \quad Let $T^*$ represent the true event time, which may or may not be observed for subjects indexed by $i = 1,2, \dots, N$. The censoring time for subject $i$ is denoted by $C_i$, and the observed follow-up time is given by $T_i = \min(T_i^*, C_i)$. Define the censoring indicator as $\delta_i = I(T_i^* \leq C_i)$, where $\delta_i = 1$ if the event is observed and $\delta_i = 0$ otherwise, indicating censoring. The time-to-event distribution within each latent class can be modeled using a Cox proportional hazards framework, incorporating a class-specific baseline hazard function and corresponding parameters:
	$$\begin{aligned}
		\lambda_{i}(t|R_{ij}=k)=\lambda_{0k}(t)\exp(\bm{X}_{3i}^T(t)\bm{\omega}_{k}+\upsilon_i)
	\end{aligned}  
	\eqno{(3)}
	$$
	In this submodel, the baseline hazard function for class $k$ is represented by $\lambda_{0k}(t)$. The covariate vector for subject $i$ at time $t$ is denoted as $\bm{X}_{3i}(t)$, while $\bm{\omega}_{k}$ represents the corresponding parameter vector for class $k$. The term $\upsilon_i$ accounts for the random effects in the time-to-event model.	For simplicity, we adopt a Gompertz baseline hazard function, expressed as $\lambda_{0k}(t) = \lambda_{0k} \exp(\gamma_k t)$, where $\lambda_{0k}(t)$ represents the parametric baseline hazard (Austin, 2012). Different from the survival submodel with the shared random effects (Miao and Charalambous, 2022),  here we get a new form of survival submodel with random effects $\upsilon$, which shares the same distribution with longitudinal submodel.\\
	
	\subsection{Variance-covariance regression submodel }
	\indent \quad Motivated by Huang et al. (2011) and He $\&$ Luo (2016), we model the variance-covariance matrix to get more information through regression submodel, which may help to get unbiased estimates for all the parameters. We model the association between longitudinal measurements and survival outcomes by the assumption of the multivariate normal distribution for two random effects $U_i$ and $\upsilon_i$:\\
	$$\begin{aligned}
		\bm{W}_i=
		\begin{pmatrix} 
			
			\bm{U}_{i}\\
			\upsilon_{i} 
		\end{pmatrix} \sim N_{(q+1)} \begin{pmatrix} 
			\begin{pmatrix} \bm{0}\\0 \end{pmatrix},\Sigma_i= \begin{pmatrix}
				\Sigma_{u_i} \qquad \Sigma_{u\upsilon_i}\\
				\Sigma_{u\upsilon_i}^T \qquad \sigma_{\upsilon_i}^2
		\end{pmatrix} \end{pmatrix} 
	\end{aligned}  
	\eqno{(4)}
	$$
	
	\quad Denote the random variables $e_{ig}=W_{ig}-\sum_{l=1}^{g-1}\phi_{igl}W_{il}$, \ $g=1,2, \dots, q+1$, $l=1,2, \ldots, g-1$. $cov(e_{ig},e_{ik})=0$ if $g \ne k$ ($1 \le j,k \le q+1, i=1,2, \dots, N$).
	Similar to Pourahmadi (1999) and Huang et al. (2011), we model the covariance matrices $\Sigma_i$ by a modified Cholesky decomposition $T_i\Sigma_iT_i^T=D_i$, where $D_i$ is a diagonal matrix with positive entries, the prediction error variances $d_{ig}^2=var(e_{ig})$ ($g=1,2, \dots q+1$); $T_i$ is the unique unit lower triangular matrix with $1$'s as diagonal entries and $-\phi_{igl}$ as its ($g,l$)th entry. The modified  Cholesky decomposition can ensure the positive definiteness of matrix $\Sigma_i$, without the need to impose constraints on any of the parameters. \\
	\indent \quad We use  linear and log linear models for the generalised autoregressive parameters $\phi_{igl}$ and innovation variances of $d_{ig}^2$, accordingly, 
	i.e. :\\
	$$ \left\{
	\begin{aligned}
		&\phi_{igl}=\bm{A}_{igl}^T \ \bm{ \alpha}_1  \\
		&logd_{ig}^2=\bm{B}_{ig}^T \ \bm{\alpha}_2\\
	\end{aligned}
	\right.
	\eqno{(5)}
	$$
where $\bm{A}_{igl}$, $\bm{B}_{ig}$ are covariates ($i=1,2,\ldots N$; $g=1,2,\ldots q+1$; $l=1,2,\ldots g-1$) and  $\bm{\alpha}_1$, $\bm{\alpha}_2$ are low-dimensional parameter vectors. The association   between the longitudinal and survival outcomes is partly described  by $\Sigma_{u\upsilon_i}$ and we can  assess it's strength  by testing the hypothesis $\Sigma_{u\upsilon_i}=\bm{0}$. \\
	\indent \quad We can see that our general  JLCM uses both latent classes and heterogeneous random effects to link the  longitudinal and survival outcomes. The addition of the variance-covariance regression submodels, allows us to assess any effects of covariates on the association between the two processes. As previously, we employ an MCMC approach for  estimation, which is described below.\\
	
	\subsection{Estimation}
	\subsubsection{The likelihood }	
	\indent	\quad   Let $\bm{\Psi}$ represent the collection of model parameters, defined as  
	\[
	\bm{\Psi} = (\bm{\xi}, \bm{\beta}, \bm{\omega}, \bm{\gamma}, \bm{\tau}, \bm{\lambda}, \bm{\alpha}_1, \bm{\alpha}_2),
	\]
	where  
	\[
	\bm{\xi} = (\bm{\xi}_{1}, \dots, \bm{\xi}_{K}), \quad  
	\bm{\beta} = (\bm{\beta}_{1}, \dots, \bm{\beta}_{K}), \quad  
	\bm{\omega} = (\bm{\omega}_{1}, \dots, \bm{\omega}_{K}), \quad  
	\bm{\gamma} = (\bm{\gamma}_{1}, \dots, \bm{\gamma}_{K}),
	\]
	\[
	\bm{\tau} = ({\tau}_{1}, \dots, {\tau}_K), \quad  
	\bm{\lambda} = ({\lambda}_{01}, \dots, \lambda_{0K}), \quad  
	\bm{R}_{i} = (R_{i1}, \ldots, R_{im_i}),
	\]
	\[
	\bm{\alpha}_1 = (\alpha_{11}, \dots, \alpha_{1n_{\alpha_1}}), \quad  
	\bm{\alpha}_2 = (\alpha_{21}, \dots, \alpha_{2n_{\alpha_2}}).
	\]
	Here, $n_{\alpha_1}$ denotes the dimension of $\bm{A}_{igl}$, while $n_{\alpha_2}$ represents the dimension of $\bm{B}_{ig}$. We assume that the observed data $(\bm{Y}, \bm{T}, \bm{\delta})$ are independent given the latent class indicator $\bm{R}$ and the random effects $\bm{W} = (\bm{U}, \upsilon)$. Additionally, we assume that $(\bm{T}, \bm{\delta})$ depends only on $R_{im_i}$. Under these assumptions, the likelihood contribution for the $i$th patient is expressed as:\\
		$$\begin{aligned}
			&L_i(\bm{\Psi}|\bm{R}_{i}=k,\bm{y}_{i},T_i,\delta_i,\bm{W}_i)\\
			&=f(\bm{y}_{i}|\bm{R}_{i}=k,\bm{W}_i,\bm{\Psi})f(T_i|\delta_i, {R}_{im_i}=k,\bm{W}_i,\bm{\Psi}) P(\bm{R}_{i}=k|\bm{\Psi}) f(\bm{W}_i|\bm{\Psi}) 
		\end{aligned}  
		$$
Thus, we derive the expression for the joint likelihood function:\\
	$$\begin{aligned}
		L_{}(\bm{\Psi}|\bm{R}, \bm{Y}, \bm{T},\bm{\delta},\bm{W})
		=\prod_{i=1}^{N}f(\bm{W_i}|\bm{\Psi})\prod_{j=1}^{m_i}\prod_{k=1}^{K}P_{ijk} 
	\end{aligned}  
	\eqno{(6)}
	$$
	where 
	\begin{align*}
		P_{ijk}&=\left\{
		\begin{array}{ll}
			\{P(R_{ij}=k|\bm{\Psi})f(y_{ij}|R_{ij}=k, \bm{W}_i,\bm{\Psi})\}^{I(R_{ij}=k)}
			&
			\quad\text{for  } j \neq m_i \\
			\\
			\{P(R_{im_i}=k|\bm{\Psi})f(y_{im_i}|R_{im_i}=k, \bm{W}_i,\bm{\Psi})f(T_i|\delta_i, {R}_{im_i}=k,\bm{W}_i,\bm{\Psi})\}^{I(R_{im_i}=k)}
			&
			\quad\text{for  } j = m_i ,
		\end{array}
		\right. 
	\end{align*}
$$	f(y_{ij}|R_{ij}=k, \bm{W}_i)= (2\pi\tau_k )^{-\frac{1}{2}}\exp\{-{1}/{(2\tau_k)}(y_{ij}- \bm{X}_{2i}^T(t_{ij})\bm{\beta}_{k}-\bm{Z}_{i}^T(t_{ij})(\bm{U}_i)^2\},$$
	$$f(T_i|\delta_i, {R}_{im_i}=k,\bm{W}_i,\bm{\Psi}) = \lambda_{ik}(t|R_{im_i}=k, \bm{W}_i,\bm{\Psi})^{\delta_i} H_{ik}(t_{}|R_{im_i}=k, \bm{W}_i,\bm{\Psi})$$
	and 

	$$\begin{aligned}
		P(R_{ij}=k)=\pi_{ijk}=\frac{\exp(\bm{X}_{1i}^T(t_{ij})\bm{\xi}_{k})}{\sum\limits_{s=1}^{K}\exp(\bm{X}_{1i}^T(t_{ij})\bm{\xi}_{s})} \  \  \ \forall~~k=1, \ldots, K
	\end{aligned}  
	$$
	\setcounter{equation}{6}
	$$\begin{aligned}
		\lambda_{ik}(t|R_{ij}=k)^{\delta_i}=
		(\lambda_{0k}\exp(\gamma_kt)\exp(\bm{X}_{3i}^T(t)\bm{\omega}_{k}+\bm{U}_i))^{I(T_i^*\le C_i)}
	\end{aligned} 
	$$
	$$\begin{aligned}
		H_{ik}(t_{})=\int_{0}^{t_{}}
		\lambda_{0k}\exp(\gamma_ku)\exp(\bm{X}_{3i}^T(u)\bm{\omega}_{k}+\upsilon_i)du\\
	\end{aligned}
	\eqno{(7)}
	$$
\indent\quad Since $e_{ig}=W_{ig}-\sum_{l=1}^{g-1}\phi_{i,gl}W_{il}$,  $\phi_{igl}=\bm{A}_{igl}^T\bm{\alpha}_1 $ and $logd_{ig}^2=\bm{B}_{ig}^T\bm{\alpha}_2$ ($i=1,2,\ldots, N; \ g=1,2,\ldots, q+1; \ l=1,2,\ldots, g-1$), we can get $e_{ig}  \sim N(0,d_{ig}^2)$. So we can get the pdf of $ f(\bm{W}_i|\bm{\Psi})$ as follows:
$$\begin{aligned}
	f(\bm{W}_i|\bm{\Psi})=&\prod_{g=1}^{q+1}\frac{1}{\sqrt{2\pi d_{ig}^2}}\exp\Bigg\{-\frac{1}{2}\sum_{g=1}^{q+1}\dfrac{(W_{ig}-\sum\limits_{l=1}^{g-1}\phi_{igl}W_{il})^2}{d_{ig}^2}\Bigg\}\\
	=&\frac{1}{\sqrt{2\pi}}\exp\Bigg\{-\frac{1}{2}\sum_{g=1}^{q+1}\Big[logd_{ig}^2+(W_{ig}-\sum\limits_{l=1}^{g-1}\phi_{igl}W_{il})^2 \cdotp {d_{ig}^2}^{-1}\Big]\Bigg\}\\
	=&\frac{1}{\sqrt{2\pi}}\exp\Bigg\{-\frac{1}{2}\sum_{g=1}^{q+1}\Big[\bm{B}_{ig}^T\bm{\alpha}_2 +(W_{ig}-\sum\limits_{l=1}^{g-1}\bm{A}_{i,gl}^T\bm{\alpha}_1 W_{il})^2 \cdotp \exp(-\bm{B}_{ig}^T\bm{\alpha}_2)\Big]\Bigg\}
\end{aligned}
\eqno{(8)}  
$$\\
		\subsubsection{MCMC sampling procedure}
	\indent \quad A Bayesian estimation approach is employed to obtain parameter estimates in the proposed JLCM. Given the likelihood in (6), the posterior distribution is expressed as:  
	\[
	\pi(\bm{\Psi}|\bm{R}, \bm{Y}, \bm{T},\bm{\delta},\bm{W}) \propto L(\bm{\Psi}|\bm{R}, \bm{Y}, \bm{T},\bm{\delta},\bm{W}) \pi(\bm{\Psi}),
	\]  
	where $\pi(\bm{\Psi})$ represents the prior distribution of the parameter vector $\bm{\Psi}$. Assuming independence among prior distributions, $\pi(\bm{\Psi})$ can be factorized as the product of individual priors for each parameter component.  
	
	Using Gibbs sampling, parameters $\bm{\beta}$, $\bm{\tau}$, and $\bm{\lambda}$ are drawn directly from their full conditional distributions. Normal priors are assigned to $\bm{\beta}$, $\bm{\omega}$, $\bm{\xi}$, $\bm{\alpha}_{1}$, $\bm{\alpha}_{2}$, $\bm{\gamma}$, and $\bm{W}$, yielding conjugate posterior distributions for $\bm{\beta}$. A gamma prior is chosen for $\bm{\lambda}$, while an inverse gamma prior is used for $\bm{\tau}$. The full conditional densities for $\bm{\beta}$, $\bm{\tau}$, and $\bm{\lambda}$ are detailed in Supplementary Appendix A.  
	
	Additionally, the latent class membership indicator $\bm{R}_{ij}$ is sampled directly from its discrete posterior distribution. The posterior probability that subject $i$ at time $j$ belongs to latent class $k$ is given by:  
	\[
	P(R_{ij}=k \ | \bm{R}, \bm{Y}, \bm{T},\bm{\delta},\bm{W}) = \frac{P_{ijk}}{P_{ij1}+ \cdots +P_{ijK}}.
	\]  
	 \\
\indent \quad For parameters $\bm{\omega}$, $\bm{\xi}$, $\bm{\alpha}_{1}$, $\bm{\alpha}_{2}$, $\bm{\gamma}$, and $\bm{W}$, the corresponding conditional distributions do not follow standard forms, making direct sampling more challenging. While the Metropolis-Hastings algorithm (Metropolis et al., 1953; Hastings, 1970) is widely used and straightforward to implement, it can suffer from convergence issues in complex models.  

To address this challenge, adaptive MCMC algorithms provide an effective alternative by dynamically adjusting key tuning parameters during the sampling process, thereby improving efficiency and convergence (Roberts \& Rosenthal, 2009). Consequently, an adaptive MCMC approach is employed in our proposed JLCM to sample parameters with non-standard distributions.  \\	
	\indent \quad Following the Adaptive Metropolis (AM) algorithm outlined by Haario et al. (2001), we define a $\ddot{d}$-dimensional target distribution $\pi(\bm{\theta})$, where $\bm{\theta} = (\bm{\omega}, \bm{\xi}, \bm{\alpha}_1, \bm{\alpha}_2, \bm{\gamma}, \bm{W})$. We then partition $\bm{\theta}$ into $\bm{\theta}_p = (\bm{\omega}, \bm{\xi}, \bm{\alpha}_1, \bm{\alpha}_2, \bm{\gamma})$, with $\ddot{d}_p$ representing the dimension of $\bm{\theta}_p$. The dimensions of $\bm{\tau}$ and $\bm{\lambda}$ are denoted by $\ddot{d}_\tau$ and $\ddot{d}_\lambda$, respectively.
	
	We assume that the elements of $\bm{\theta}_p$ follow independent standard normal priors, $N(0,1)$, with the associated density function denoted as $f_{N(0,1)}$. Similarly, the components of $\bm{\tau}$ are assigned independent vague inverse Gamma priors, $IG(0.01, 0.01)$, with the density denoted as $f_{IG(0.01, 0.01)}$. For the elements of $\bm{\lambda}$, we use independent uninformative Gamma priors, $\Gamma(0.01, 0.01)$, with density function $f_{\Gamma(0.01, 0.01)}$.
	
	The resulting posterior distribution is:
	\\
$$\begin{aligned}
	\pi(\bm{\theta},\bm{\tau}, \bm{\lambda}|\bm{R}, \bm{Y}, \bm{T},\bm{{\delta}}, \bm{\Psi}_{-(\bm{\theta}_p,\bm{\tau},\bm{\lambda})})
	&\propto\Big\{\prod_{i=1}^{N}f(\bm{W}_i|\bm{\Psi}_{-(\bm{\theta}_p,\bm{\tau},\bm{\lambda})})\prod_{j=1}^{m_i}\prod_{k=1}^{K}P_{ijk} \Big\} \\
	& \prod\nolimits_{ \ddot{d}_p}f_{N(0,1)}\prod\nolimits_{ \ddot{d}_{\tau}}f_{IG(0.01, 0.01)}\prod\nolimits_{ \ddot{d}_{\lambda}}f_{\Gamma(0.01, 0.01)}\\
\end{aligned}  
\eqno{(9)}
$$
where  $\bm{\Psi}_{-(\bm{\theta}_p, \bm{\tau}, \bm{\lambda})}^*$ represent the parameter vector excluding the components of $\bm{\theta}_p$, $\bm{\tau}$, and $\bm{\lambda}$. We can then construct an adaptive Metropolis algorithm with the proposal distribution  
\[
\pi(\bm{\theta}, \bm{\tau}, \bm{\lambda} | \bm{R}, \bm{Y}, \bm{T}, \bm{\ddot{\Delta}}, \bm{\Psi}_{-(\bm{\theta}_p, \bm{\tau}, \bm{\lambda})}^*)
\]  

which is evaluated at iteration $m$ as follows:

$$\begin{aligned}
	Q_m( (\bm{\theta}, \bm{\tau},\bm{\lambda}),\cdotp)=\left\{
	\begin{aligned}
		&N( (\bm{\theta}, \bm{\tau},\bm{\lambda}), (0.1)^2I_{\ddot{d}}/\ddot{d} ) \ \ \ for\  m \leq 2\ddot{d} \\
		&(1-\alpha_{prop})N( (\bm{\theta}, \bm{\tau},\bm{\lambda}), \sigma^2\cdotp \Sigma_m/\ddot{d})\\
		&+ \alpha_{prop}\cdotp N( (\bm{\theta}, \bm{\tau},\bm{\lambda}), (0.1)^2I_{\ddot{d}}/\ddot{d} ) \ \ \  for\  m>2\ddot{d} \\
	\end{aligned}
	\right.
\end{aligned}  
\eqno{(10)}
$$
where $\Sigma_m$ represent the current empirical estimate of the covariance structure of the target distribution, based on the results obtained so far. The term $\sigma^2$ is a fixed variance constant that we select to assess the performance of the adaptive algorithm, and ${\alpha}_{prop}$ is a small positive constant.  Gelman et al. (1997) and Roberts and Rosenthal (2001) observed that the proposal distribution $N(\bm{\theta}, (2.38)^2 \Sigma_m / \ddot{d})$ is optimal in certain large-dimensional scenarios. In our case, we found that the proposal $N(\bm{\theta}, (0.1)^2 I_{\ddot{d}} / \ddot{d})$ provides a better fit.
 \\
\indent\quad According to the standard Metropolis-Hastings algorithm (Hastings, 1970), a proposal from $(\bm{\theta}, \bm{\tau}, \bm{\lambda})$ to $(\bm{\theta}^*, \bm{\tau}^*, \bm{\lambda}^*)$ is accepted with the probability  
\[
\alpha((\bm{\theta}, \bm{\tau}, \bm{\lambda}), (\bm{\theta}^*, \bm{\tau}^*, \bm{\lambda}^*)) = \min \left( \frac{\pi((\bm{\theta}^*, \bm{\tau}^*, \bm{\lambda}^*)|\bm{R}, \bm{Y}, \bm{T}, \bm{\ddot{\Delta}}, \bm{\Psi}_{-(\bm{\theta}_p,\bm{\tau},\bm{\lambda})}^*)}{\pi((\bm{\theta}, \bm{\tau}, \bm{\lambda})|\bm{R}, \bm{Y}, \bm{T}, \bm{\ddot{\Delta}}, \bm{\Psi}_{-(\bm{\theta}_p,\bm{\tau},\bm{\lambda})}^*)}, 1 \right)
\]
where $\bm{\theta}^* = (\bm{w}^*, \bm{\delta}^*, \bm{\xi}^*, \bm{U}^*)$ represents the new parameter vector consisting of $\bm{w}, \bm{\delta}, \bm{\xi}, \bm{U}$, and $\bm{\tau}^*$ and $\bm{\lambda}^*$ are the new parameter vectors for $\bm{\tau}$ and $\bm{\lambda}$, generated as "candidate values" from the proposal distribution $Q_m((\bm{\theta}, \bm{\tau}, \bm{\lambda}), \cdot)$.  

Through the MCMC sampling process, we can also compute the posterior membership probability for model-based classification. The posterior probability of subject $i$ belonging to class $k$ at time $j$ is given by
\[
\hat{\pi}_{ijk} (\hat{\bm\Psi}) = \frac{\pi_{ijk} L_{ijk}(\hat{\bm\Psi})}{\sum_{k=1}^{K} \pi_{ijk} L_{ijk}(\hat{\bm\Psi})}
\]
where $L_{ijk}(\hat{\bm\Psi}) = f(\hat{\bm{W}}_i|\hat{\bm\Psi}) \hat{P}_{ijk}$ is computed using the posterior sample of the parameters $\bm{\Psi}$ and random effects $\bm{W}$, and $\pi_{ijk}$ is the prior probability as defined in (1). The higher the value of the posterior probability $\hat{\pi}_{ijk} (\hat{\bm\Psi})$, the more likely it is that the subject belongs to this group.
\\  

	\subsubsection{Dynamic prediction of time-to-death in survival analysis}
	\indent \quad  Similar to the procedure described in  time-varying JLCM (Miao and Charalambous, 2022), we can compute dynamic predictions of the time-to-event probabilities based on our proposed JLCM. The survival function for subject  $i$ is given by
$$\begin{aligned}
	S_i(t+\Delta t|T\geq t, y_i(t), \hat{\bm{\Psi} }, \hat{\bm{W} }_i )=\frac{ S_i\{ t+\Delta t| y_i(t), \hat{\bm{\Psi} }, \hat{\bm{W} }_i  \}  }{   S_i\{ t| y_i(t), \hat{\bm{\Psi} }, \hat{\bm{W} }_i  \}   }
\end{aligned}
\eqno{(11)}
$$
where $ \hat{\bm{\Psi} }$ and $\hat{{W} }_i $ denote the posterior means of the parameter vector $ {\bm{\Psi} }$ and random effects $W_i$, and  $y_i(t)$ contains the longitudinal measurements  up  to time $t$  to imply  survival up to this time point. $	\hat{S}_i\{ t| y_i(t),R_{ij}=k,\hat{\bm{\Psi} }, \hat{\bm{W} }_i  \}=\exp(-	\hat{H}_{ik}(t| y_i(t),R_{ij}=k, \hat{\bm{\Psi} }, \hat{\bm{W} }_i    ))$ with $\hat{H}_{ik}(\cdot)$ given in (7). 

The posterior predictive survival function $S_i$ at $t$ to predict the survival of a patient up to time $t$ is given by
$$\begin{aligned}
	\hat{S}_i\{ t|, y_i(t), \hat{\bm{\Psi} }, \hat{\bm{W} }_i \}=\sum_{k=1}^{K} \hat P(R_{im_i}=k| \hat{\bm{\Psi} })	\hat{S}_i(t| y_i(t),R_{im_i}=k,\hat{\bm{\Psi} }, \hat{\bm{W} }_i)
\end{aligned}
\eqno{(12)}
$$	
which is a  mixture  of survival distributions for patient $i$ conditional on subgroup $k$ to get  survival probabilities for each subject. Other notations and equations, such as the misclassification rate and the dynamic predictions of the time-to-event probabilities, are defined in the time-varying JLCM with shared random effects, as proposed by Miao and Charalambous (2022).\\
	\section{Simulation study of joint latent class model with time-varying probability in Bayesian approach}
		\quad  We aim to simulate 100 datasets from the proposed general JLCM, classified clearly into different groups, which allows the existence of jumping behaviours among subgroups and the assessment of any effects of covariates on the association between the longitudinal and survival processes. For each simulation, we have 6 different model settings, including the proposed general JLCM with 1 to 3 subgroups and the time-varying JLCM with 1 to 3 classifications. The Bayesian approach is adopted for estimation and the DIC criterion is used to decide the optimal number of classes. Simulation results show that the proposed general JLCM  can produce more accurate results with smaller bias values and the general JLCM outperforms the time-varying  JLCM with better classification and prediction.
	
	\subsection{Generated data information}
	\quad In order to illustrate the importance of the inclusion of covariance modelling in the joint latent class model, we conduct a simulation study to compare the performance of the general JLCM with a time-varying JLCM without covariance modelling.  The latter is similar to the time-varying JLCM proposed in time-varying JLCM (Miao and Charalambous, 2022). However, different from shared random effects, here we assume heterogeneous random effects sharing the same distribution as described earlier.  For simplicity,   in our simulation  study we just consider two latent classes
	and  use baseline covariates in  $X_{3i}$  but a more complex model with time-varying $X_{3i}$  is also possible. 
	The fixed  covariate $X_{1i}$ which is generated from a standard normal distribution, is used to capture certain  baseline characteristics of each subject and $X_{3i}$  corresponds to a  treatment indicator simulated from the bernoulli distribution  (1 if subject receives treatment  and 0 otherwise).  
	The longitudinal submodel and membership probability are in the same form as the ones in time-varying JLCM (Miao and Charalambous, 2022).  But  survival submodel is in the different form $\lambda_{ik}(t)=\lambda_{0k}\exp(\gamma_kt)\exp(w_kX_{3i}+\upsilon_i)$.  For the baseline hazard function with the form of $  h_{0k}(t)=\lambda_{0k} \exp(\gamma_k t)$, we can simulate survival times from  $T_k=\log(1-{\gamma_k \log(u)}/({\lambda_{0k} \exp(w_k X_{3}+\upsilon)}))/{\gamma_k} $  where $u\sim U(0,1)$. 
	Due to the complexity of the model, we only consider 100 Monte Carlo samples in our simulation study.  The random effects are simulated from the multivariate normal distribution with variance covariance matrix $\Sigma_i$ as in equation  (4),  with the associated regression covariates set as  $\bm{A}_{igl}=\bm{B}_{ig}=(1, X_{3i})$, where $X_{3i}$ denotes the treatments corresponding to each individual. For simplicity,  we set $q=2$ which results in a  3x3 dimensional covariance matrix for the random effects. As we have assumed that random effects will not change among different subgroups (i.e. $U_i, \upsilon_i$ will be the same across classes), $\bm{ \alpha}_1, \bm{ \alpha}_2$ should also remain the same for different groups. i.e.\\
	$$ \left\{
	\begin{aligned}
		&\phi_{igl}=\bm{A}_{igl}^T \ \bm{ \alpha}_1  \\
		&logd_{ig}^2=\bm{B}_{ig}^T \ \bm{\alpha}_2\\
	\end{aligned}
	\right.
	$$
	where $\bm{ \alpha}_1=(-0.2,-0.5)^T$, $\bm{ \alpha}_2=(0.1,0.3)^T$, $i=1,2,\ldots, N$, $g=1,2,\ldots, q+1$ and $l=1,2,\ldots, g-1$.\\
\indent \quad 
	The other submodels vary across classes and are given as follows:\\

	\textbf{Class 1}:\\
	Latent class: $p_{ij1}=\frac{\exp(0.01X_{1i}+0.2Time_{ij})}{(\exp(0.01X_{1i}+0.2Time_{ij})+\exp(Time_{ij}))}$ \\
	Longitudinal submodel: $y_{ij1}=2X_{1i}+1.5Time_{ij}+U_{1i}+U_{2i}Time_{ij}+\epsilon_{ij1} $\\ 
	We specify  $\epsilon_{ij1} \stackrel{i.i.d} \sim N(0,{0.1})$ (0 mean and the variance is 0.1) \\
	Survival submodels: $\lambda_{i1}(t)=0.2\exp(0.2t)\exp(0.5X_{3i}+\upsilon_i) \  \ \ for \ t \le T_i$\\
	
	\textbf{Class 2}:\\
	Latent class: $p_{ij2}=\frac{\exp(Time_{ij})}{\exp(0.01X_{1i}+0.2Time_{ij})+(\exp(Time_{ij}))}$ \\
	Longitudinal submodel: $y_{ij2}=4X_{1i}+3Time_{ij}+U_{1i}+U_{2i}Time_{ij}+\epsilon_{ij2} $\\ 
	We specify $\epsilon_{ij2} \stackrel{i.i.d} \sim N(0,{0.5})$ \\
	Survival submodels: $\lambda_{i2}(t)=0.1exp(-0.2t)\exp(0.8X_{3i}+\upsilon_i)  \  \ \ for \ t \le T_i$\\
	
	The true values above in the simulation studies are the most sensible values for the general JLCM, chosen from different tries of value settings.
	
	\subsection{Simulation results }
		\quad  We have the same settings as the simulation setup in time-varying JLCM (Miao and Charalambous, 2022) for total numbers of individuals and records. For  each simulation run, we set $N$ as the total number of subjects and $m_i$ as the number of repeated longitudinal measurements for each subject $i$. We illustrate the features of the datasets created under our general JLCM with shared random effects and covariance modelling, by focusing on one simulated dataset. 
		Firstly we focus on the basic JLCM which does not allow the group membership probability to change with time.	The plot of longitudinal trajectories is shown in the left of Figure 1. \\
			\begin{figure}[H]
			\begin{center}
				\includegraphics[width=1.05\textwidth, height=0.37\textwidth]{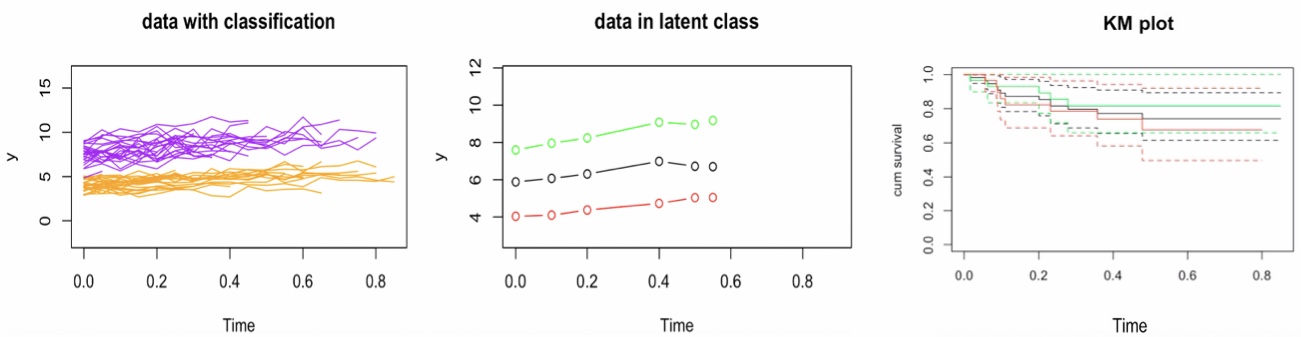}
				\center{Fig. 1.   Left: Longitudinal trajectories with certain classification (orange: class 1/low level; purple: class 2/high level); Middle: Mean longitudinal trajectories; Right: K-M plot (black: no classification; red: class 1/low level; green:
					class 2/high level)}
			\end{center}
		\end{figure}
		\indent \quad 	The mean longitudinal trace and Kaplan-Meier plot, based on the basic JLCM without time-varying changes or covariance modelling, are also presented in Figure 1. It is clear that incorporating classification is crucial to differentiate between the two distinct subgroups in this dataset. However, this static classification approach does not provide insight into how individuals may shift group membership over time.\\
			\indent \quad Tables 1 and 2 summarize the results of our simulation study, which involved 100 simulated datasets and 5000 MCMC iterations, with 2000 iterations used for burn-in. The optimal number of classes for each model was determined using the Deviance Information Criterion (DIC). From Table 1, we observe strong evidence supporting the correct model, which is the proposed general JLCM with covariance modelling and 2 classes, as indicated by the DIC. In 40\% of cases, when the time-varying JLCM was selected as the optimal model (according to the error rate), we found that a model with 2 classes was preferred. This is not surprising, as the time-varying JLCM is a special case of the general JLCM, and we would expect the correct number of classes to be identifiable regardless of whether covariance modelling is included. Additionally, we note that when the simpler time-varying JLCM was selected as the optimal model, the error rates were very similar to those observed in the general JLCM. 	In Table 2, we observe that parameter estimation is accurate, with minimal bias. However, the posterior standard deviations for most parameters are inflated, likely due to the use of only 3000 MCMC iterations (after burn-in) for posterior inference. As a result, the credible intervals for some smaller effects, such as $\xi_{11}$, $\xi_{21}$, and $\alpha_{21}$, include 0.
			 \\
	{ \centerline{ Table 1:	{ DIC and error rate over 100 simulated datasets and 3000 MCMC iterations}
	}}
\setlength{\tabcolsep}{4.4mm}{
	\begin{longtable}{llllllccc}
		\hline			
		\multicolumn{6}{c}{the general JLCM  	 }	 
		&\multicolumn{1}{c}{time-varing JLCM} 
		\\ 
		\hline
		Model      & $K$=1   & $K$=2 & $K$=3  \  & $K$=1 & $K$=2 & $K$=3 & $K$=4\\ 	\hline	
		Model selection via DIC  &  \ \ 0.01& \ \ 0.99&  \ \ 0  \ \ \  \   &\ \ 0 &  \ \ 0 & 0&0    \\
		Model selection via error rate   & \ \ 0  & \ \ 0.6  & \ \ 0  \ \ \  \  & \ \ 0  &\ \ 0.4   &  0 &0 \\
		\hline	
	\end{longtable} }

		{	\centerline{ Table 2: 	{	Estimation results over 100 simulated datasets and 3000 MCMC iterations, based on the general JLCM with $K$=2
			}
		}
	}
	
	\setlength{\tabcolsep}{3.3mm}{
		\begin{longtable}{llllllcc}
			\hline
			\multicolumn{6}{c}{\ \ \ \ \ \ \ \ \ \ \ \ \ \ \ \ \ \ \ \ \ \ \ \ \ \ \ \ \ \ \ \ \ \ \ \ \ \ \ \ \ \  time-varying JLCM with 	$K$=2 	 }	 \\ \hline 
			\multicolumn{4}{c}{Class 1}	 
			&\multicolumn{4}{c}{Class 2} 
			\\ 
			\hline
			Parameter           & Estimate (Bias) &sd  & CI(89\%)& 	Parameter  &Estimate (Bias) &sd  & CI(89\%)\\ \hline 
			
			$\beta_{11}$         &  2.0952  (0.0952)  & 0.3310&[2.03, 2.20]&	$\beta_{21}$  & 3.9029(-0.0971)   &  0.2184& [3.81, 3.98]\\
			$\beta_{12}$         &   1.5834  (0.0834) &  0.6463&  [1.24, 2.02]&$\beta_{22}$ &2.9065 (-0.0935)& 0.4587 & [2.51, 3.23] \\
			$\tau_1$          & 0.1067   (0.0067)  &  0.0715&   [0.08, 0.16]&$\tau_2$         & 0.4419  (-0.0581) &0.2776  & [0.34, 0.55] \\
			$\lambda_{1}$   &  0.1679 (-0.0321) &  0.1108&  [0.13, 0.21]&$\lambda_2$       & 0.0835 (-0.0165) &  0.0585  &  [0.05, 0.13]\\
				$\gamma_{1}$   &  0.1846(-0.0154) &  0.1912 &    [0.12, 0.26]  & $\gamma_{2}$   &  -0.1912 (0.0088) &  0.2161&   [-0.28, -0.13] \\
			$	w_1    $   & 0.4490 (-0.0510)  & 0.3351 &[0.35, 0.58]&	$w_2  $   & 0.7246 (-0.0754)  &   0.5036& [0.57, 0.91] \\	
				$\xi_{11}$           &0.0112(0.0012) &  0.1776 &  [-0.06, 0.08] &$\xi_{21}$                & -0.0079 (-0.0079) &  0.1606 & [-0.10, 0.06]\\ 
			$\xi_{12}$          & 0.1738 (-0.0262) & 0.2243   &  [0.10, 0.25]&$\xi_{22}$            & 0.9017 (-0.0983) & 0.5934  &   [0.71, 1.09]\\	
		\hline
		\multicolumn{8}{c}{\ \ \ \ \ \ \ Same value across two classes	 }	 \\ \hline 
		${\alpha}_{11}$   &-0.1727 (0.0232)  &   0.2070 &     [-0.26, -0.11]  &
		${\alpha}_{12} $  &-0.4540(0.0460) &  0.3526 &  [-0.58, -0.35]  \\
		${\alpha}_{21} $  &0.0855 (-0.0145) &0.1966&  [-0.01, 0.18] &
		${\alpha}_{22}$  & 0.2655 (-0.0345)   &  0.2633 &   [0.16, 0.39]  \\ \hline
	\end{longtable} }
	
		\subsection{Dynamic predictions of time to death}
			\indent \quad	We further evaluate the performance of our proposed general model by comparing the expected survival outcomes of individuals with those obtained using the time-varying JLCM. The survival predictions are calculated using equations (11) and (12), and the results for eight selected subjects are presented in Figures 2 and 3. These plots display the combined information on longitudinal trajectories and survival probabilities for each chosen subject. The green dotted line represents the dynamic predictions from the proposed general JLCM with 2 classes (the true model), while the blue dotted line corresponds to the survival probabilities from the time-varying JLCM with 2 subgroups (the best alternative model, selected based on the smallest error rate). The vertical dashed line marks the time to event (in red) or censoring (in purple).\\
			\indent \quad  For subject 25 in Figure 2, we observe that at the time of the event, the survival probability from the proposed model is slightly lower than that from the basic JLCM, with values of 0.9792 and 0.9860, 
			respectively. For subjects with stable trajectories (e.g., subjects 15, 19, and 35), the proposed model with covariance modelling yields marginally lower but similar survival probabilities compared to the time-varying JLCM. In contrast, for individuals exhibiting abrupt changes, as shown in Figure 3, the general JLCM is more conservative, providing lower predictive survival probabilities than the time-varying JLCM.\\
			\indent \quad 	In other words, the general JLCM is more responsive to changes in group membership. For example, for subject 8, the survival probability is more sensitive to the drop in the response at the final two measurements. This is not unexpected, as subjects exhibiting abrupt changes tend to have more volatile trajectories, and failing to account for this variability may affect their survival predictions. The dynamic prediction plots for six of these subjects are presented in Figure C1 of Appendix C.\\
			
			\begin{figure}[H]
				\begin{center}
					\includegraphics[width=0.4\textwidth, height=0.26\textwidth]{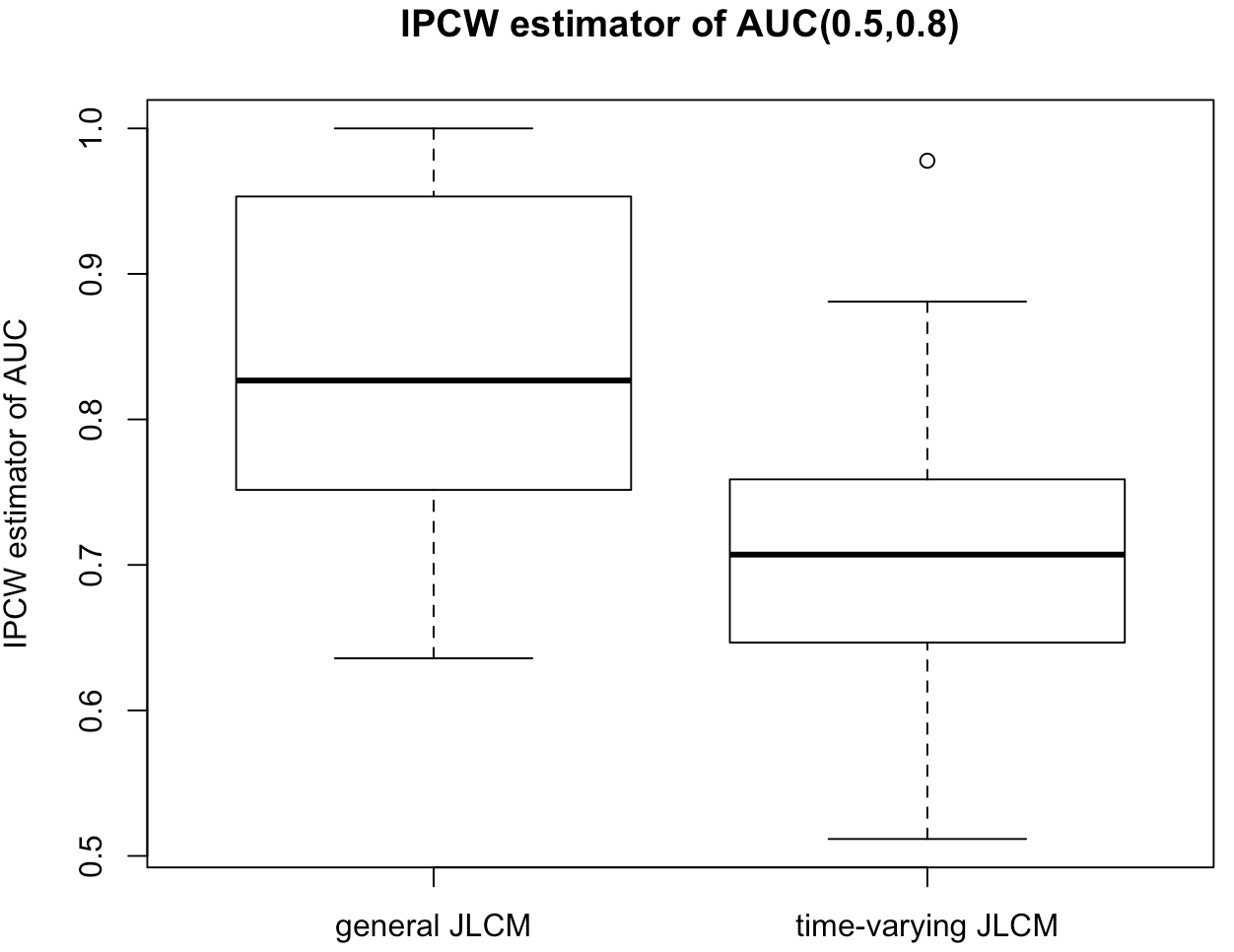}
					\center{Fig. 4 Boxplots of the IPCW estimator of AUC when assuming the general and time-varying JLCM at time point  $t=0.5$ with $\Delta t=0.3$}
				\end{center}
			\end{figure}
			
		\begin{figure}[H]
			\begin{center}
				\includegraphics[scale=0.35]{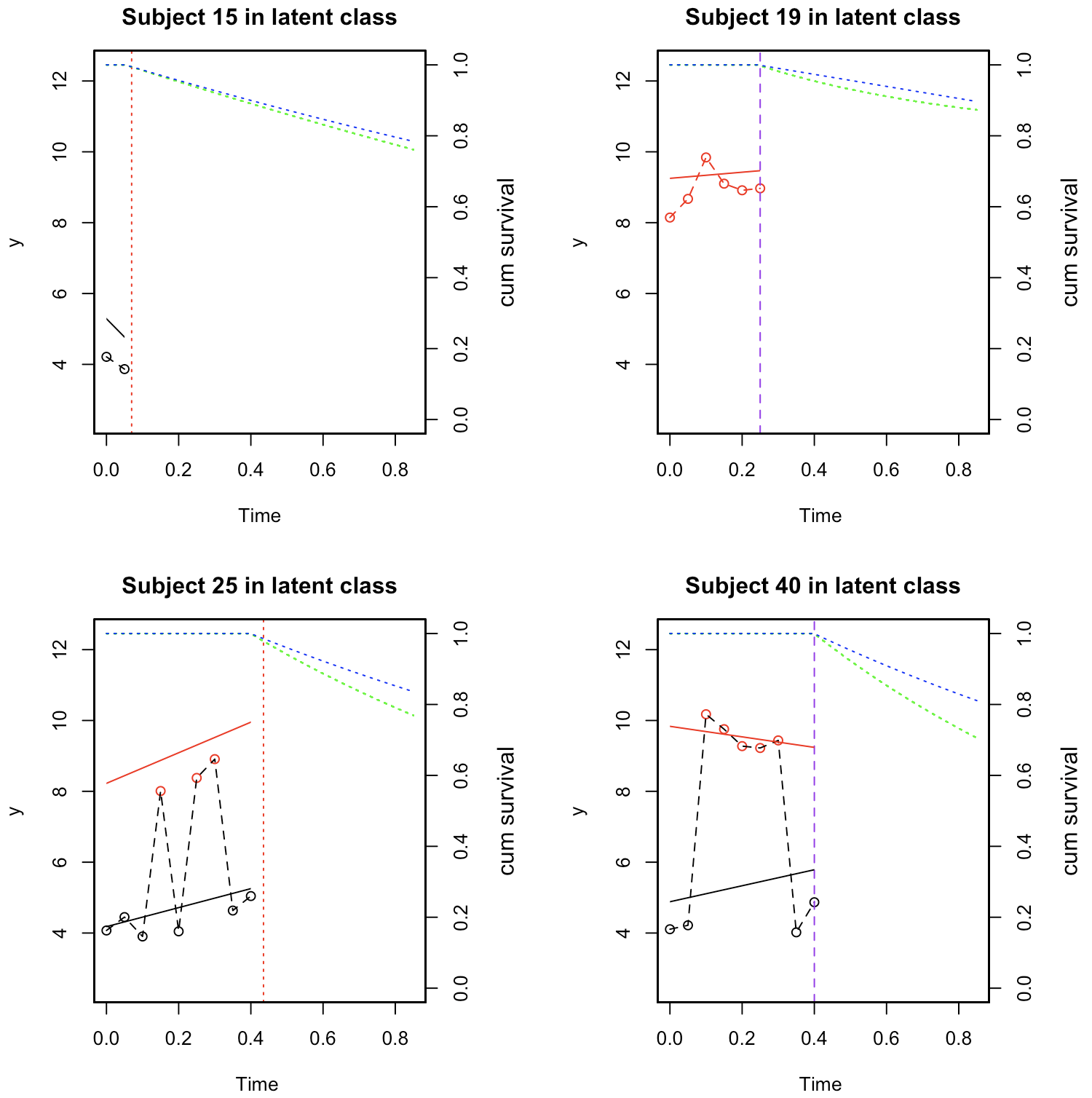}
			\end{center}
			\center{ Fig. 2 Prediction of simulated data for four cases from both proposed and basic JLCM (black point: class 1/low level; red point: class 2/high level; black dashed  line: longitudinal trajectory which stayed in class 1; red  dashed  line: longitudinal trajectory which stayed in class 2; black full line: fitted trajectory which stayed in class 1; red  full line: fitted trajectory which stayed in class 2;  green dotted line: survival probability from proposed JLCM; blue dotted line: survival probability from basic JLCM; purple vertival dashed line: subject censored; red  vertical dotted line: time to event happened)
			}
		\end{figure}

	\begin{figure}[H]
		\begin{center}
			\includegraphics[scale=0.35]{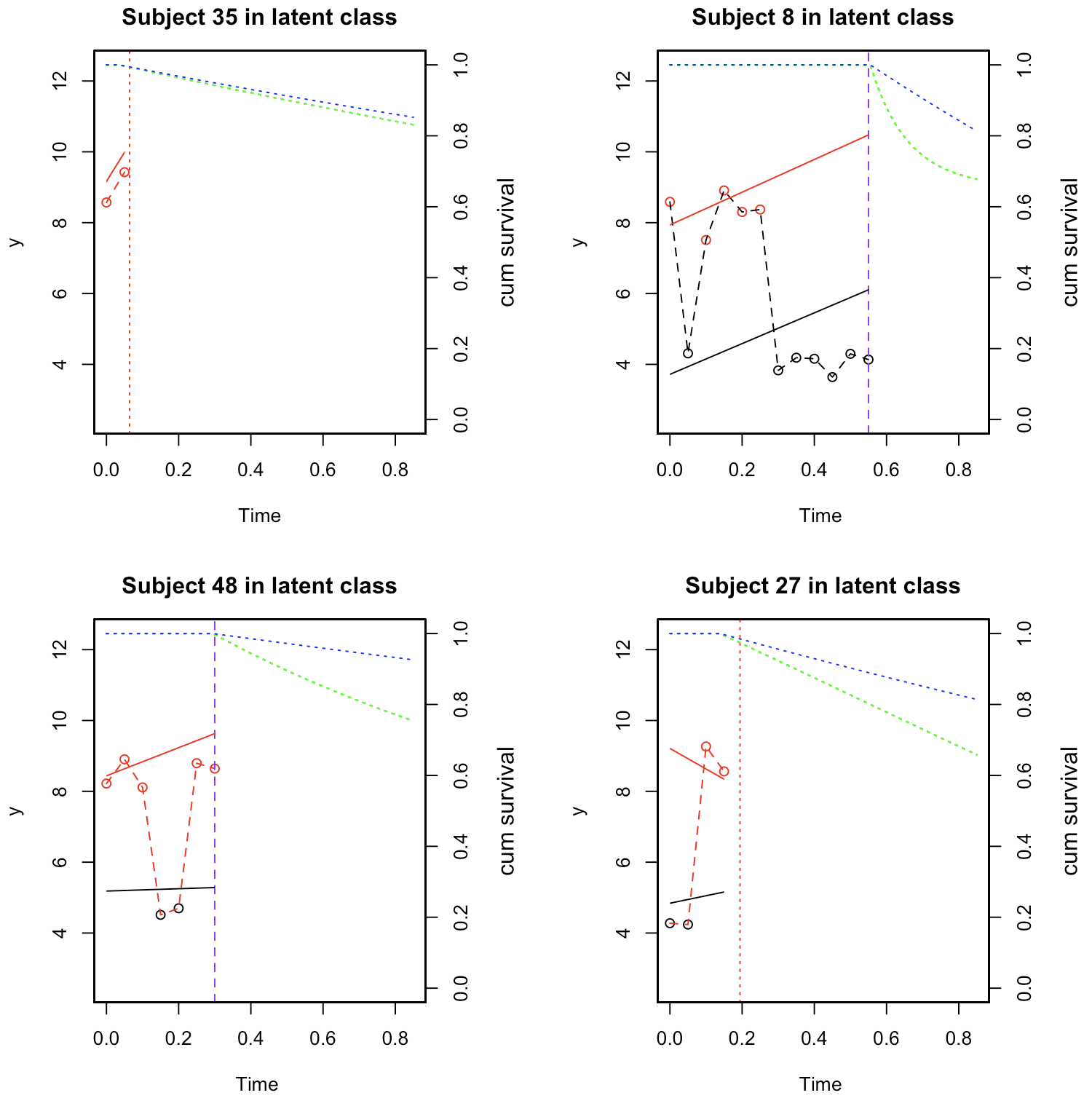}
		\end{center}
		\center{ Fig. 3 Prediction of simulated data for four cases from both proposed and basic JLCM (black point: class 1/low level; red point: class 2/high level; black dashed  line: longitudinal trajectory which stayed in class 1; red  dashed  line: longitudinal trajectory which stayed in class 2; black full line: fitted trajectory which stayed in class 1; red  full line: fitted trajectory which stayed in class 2;  green dotted line: survival probability from proposed JLCM; blue dotted line: survival probability from basic JLCM; purple vertival dashed line: subject censored; red  vertical dotted line: time to event happened)
		}
	\end{figure}
		
	\indent \quad 	To assess the prediction accuracy of the general and time-varying JLCM, we calculate the IPCW estimator of the AUC within the time interval $[0.5, 0.8)$ at $t = 5$ with $\Delta t = 0.3$ over 100 simulations, following the AUC equation from JLCM (Miao and Charalambous, 2022). The boxplots of the IPCW estimator of AUC, denoted as $\widehat{AUC}(0.5, 0.8)$, are presented in Figure 4.\\

\indent \quad 	In Figure 4, we can obtain higher IPCW estimators from the general JLCM, which indicates that the general JLCM outperforms the time-varying JLCM due to better predictions.\\

	\section{Real data application}

\subsection{Data information}
	\indent\quad  The AIDS dataset (Goldman et al., 1996) comprises both longitudinal and survival data from a randomized clinical trial aimed at comparing the efficacy and safety of two antiretroviral drugs in patients who had either failed or were intolerant to zidovudine (AZT) therapy. The dataset consists of 1,408 observations across 9 variables: $patient$, $Time$, $death$, $CD4$, $obstime$, $drug$, $gender$, $prevOI$, and $AZT$. There are a total of 467 distinct patients, each identified by a unique patient identifier. The variable $Time$ represents the time to death or censoring, while $death$ is a binary indicator (0 for censoring, 1 for death). 
	
	The $CD4$ variable represents the CD4 cell count in the blood, measured at study entry and at 2, 6, 12, and 18 months. CD4 cells are white blood cells crucial for fighting infections and are used to assess the health of the immune system in HIV-infected individuals. The variable $obstime$ indicates the time points at which CD4 cell counts were recorded. 
	
	The $drug$ variable is a factor with two levels: ddC (zalcitabine) and ddI (didanosine). $gender$ is also a factor with two levels: female and male. The variable $prevOI$ is a factor with levels AIDS (denoting previous opportunistic infections, specifically AIDS diagnosis at study entry) and noAIDS (denoting no previous infection). Finally, $AZT$ is a factor with two levels: intolerance and failure, indicating whether the patient experienced intolerance or failure to AZT therapy.	\\
	\indent\quad For the real data analysis, we select the AIDS dataset and fit both the proposed general JLCM and the time-varying JLCM, considering models with 1, 2, or 3 latent classes.\\
	
	\subsection{Estimation results and do the comparison with results from time-varying JLCM}
	\indent \quad We choose AIDS drug trial data to do the comparison between the proposed  general and time-varying JLCM. AIDS dataset is firstly to be applied in joint modelling research. Six different model settings are chosen as the simulation modelling setting, including the proposed general JLCM with subgroup numbers from 1 to 3 and the time-varying JLCM with classification settings from 1 to 3. A baysian approach is used through MCMC to implement the these models.\\
		\indent \quad For the AIDS data, we set the low CD4 level as class 1 and high CD4 level as class 2 (same settings as simulation study). 
		To fit the general JLCM for the AIDS data, the longitudinal submodel  is as follows:\\
		$$\begin{aligned}
		CD4_{ijk}=&\beta_{1k}+\beta_{2k}Time_{ij}+\beta_{2k}gender_{ij}+\beta_{4k}prevOI_{ij}+\beta_{5k}AZT_{ij}
		+U_{1i}+U_{2i}Time_{ij}+\epsilon_{ijk}\\
	\end{aligned}
	$$\\
	\indent \quad	 For the hazards submodel, we consider:\\
	$$\begin{aligned}
		\lambda_{ik}(t)=\lambda_{0k}\exp(\gamma_kt)\exp(w_{k1}gender_i+w_{k2}prevOI_i+w_{k3}AZT_i+w_{k4}drug_i+\upsilon_i )\\
		for \ t \le T_i\\
	\end{aligned}
	$$\\
		\indent \quad	The random effects follow a  multivariate normal distribution with  mean zero and the variace covariance $\Sigma_i$, as follows \\
	
	$$\begin{pmatrix} 
		\bm{U}_{i}\\
		\upsilon_{i} 
	\end{pmatrix} \sim N_{(q+1)} \begin{pmatrix} 
		\begin{pmatrix} \bm{0}\\0 \end{pmatrix},\Sigma_i=\begin{pmatrix}
			\Sigma_{	{u}_i} \qquad \Sigma_{	{u}\upsilon_i}\\
			\Sigma_{	{u}\upsilon_i} \qquad \sigma_{\upsilon_i}^2
	\end{pmatrix} \end{pmatrix} $$\\

		\indent \quad	  For the regression submodel, we set that	
		$$ \left\{
		\begin{aligned}
			&\phi_{igl}=\bm{A}_{igl}^T \ \bm{ \alpha}_1  \ \ \ \ \  i=1,2,\ldots, N; \ \ \  g=1,2,\ldots, q+1 \\
			&logd_{ig}^2=\bm{B}_{ig}^T \ \bm{\alpha}_2 \ \ \ l=1,2,\ldots, g-1\\
		\end{aligned}
		\right.
		$$\\
where we specifically choose $\bm{ A}_{igl}=\bm{ B}_{ig}=(drug_i, AZT_i)^T$, such that a heterogeneous  covariance matrix is allowed for different treatment groups. \\
	\indent \quad	Finally, for the  latent class sumodel, we specify\\
	$$\begin{aligned}
		p_{ijk}=\frac{\exp(\xi_{1k}+\xi_{2k}Time_{ij})}{\sum\limits_{s=1}^{2}\exp(\bm{X}_{1i}^T(t_{ij})\bm{\xi}_{s})}\  \  \ \forall~~k=1, \ldots, K
	\end{aligned}
	$$\\
	\indent \quad	where	 $Time_{ij}$ denotes the measurement time  for the  patient $i$ at time $j$. $gender_{ij}$ is a factor with levels female (0) and male (1).  $prevOI_{ij}$ is the a factor with levels AIDS  denoting previous opportunistic infection (AIDS  diagnosis) at study entry, and no AIDS  denoting no previous infection. 
	$AZT_{ij}$ is the factor with levels intolerance and failure denoting AZT intolerance and AZT failure, respectively. 
	$U_{1i}$ is the random intercept and $U_{2i}$ is treated as the random slope for time in this linear mixed-effects model 
	with the bivariate normal distribution $	\bm{U_{i}} \sim N_{(2)}  (  \bm{0},\Sigma_{ui}   )$			
	and $\epsilon_{ijk}$ is the measurement errors which follow a normal distribution: $\epsilon_{ijk} \stackrel{i.i.d} \sim N(0,\tau_k)$. \\	
	\indent \quad	For comparison,  we also fit a time-varying JLCM with shared   distribution. The longitudinal submodel, survival submodel and latent class submodel  are set the same as the fitted general JLCM above.\\
	
	{ \centerline	{Table 3: 	{DIC table of AIDS data for both general and time-varying JLCMs in Bayeian approach}
	}
}

\begin{longtable}{llllllccc}
	\hline			
	\multicolumn{4}{c}{\ \ \ \ \ \ \ \ \ \ \ the general JLCM 	 }	 
	&\multicolumn{2}{c}{\ \ \ \ \ \ \ \ \ \ \ time-varying JLCM} 
	\\ 	\hline
	Model      & \ \ \ \ \ \ \ $K$=1   & $K$=2 & $K$=3 \ \ \ \ \ \ \   \ \ \ \ \ & \ \ \ \ \ $K$=1  \ \ \    \ \ \   &\ \ \ \ $K$=2 &$K$=3 \\ 	\hline	
	DIC   & \ \ \ \ \ \ \ 537.2458\ \ \  &  483.5541\ \ \ \ \ \ & 572.3161  \  &  \ \ \ \ \ 894.5011  \ \ \   & \ \ \ \  864.4287 \ \ \ \ \ \ & 902.6324 \\
	\hline			
\end{longtable}	
	
		\indent \quad As shown in Table 3, the proposed general JLCM with 2 classes is identified as the optimal model. For the time-varying JLCM, the optimal model is also the one with 2 classes, which aligns with the results from our simulation study. Further details on the estimation results from the time-varying JLCM can be found in Appendix B.\\	
		\indent \quad Following 2000 iterations for burn-in, the posterior means, standard deviations, and 89\% credible intervals for each parameter in both the time-varying and general JLCMs are presented in Table B1 (Appendix B) and Table 4, respectively.\\

		{ \centerline	{Table 4:	{Estimation results for the analysis of the AIDS data based on the general JLCM with $K$=2
				}
			}
		}	
		\setlength{\tabcolsep}{3.3mm}{
			\begin{longtable}{llllllcc}
				\hline
				\multicolumn{4}{c}{Class 1}	 
				&\multicolumn{4}{c}{Class 2} 
				\\ 
				\hline
				Parameter           & Estimate  &sd  & CI(89\% )& 	Parameter  &Estimate &sd  & CI(89\% )\\ \hline 
				$\beta_{11}$ &   8.2885     &        1.5744        &   [5.79, 10.72]	
				&	$\beta_{21}$ & 13.1232  & 1.4409&  [10.66, 15.15]	\\
				$\beta_{12}$   &  -0.1405     &          0.0683    &    [-0.26, -0.06]	
				&	$\beta_{22}$   
				&  -0.0738 &  0.0580& [-0.15, 0.02]	   	 \\
				$\beta_{13}$&    -0.1776  &         0.8046       &   [-1.53, 0.95]		&$\beta_{23}$  				      
				&   -0.1365    &       0.7517      &   [-1.33, 1.05]	     \\
				$\beta_{14}$ &   -2.6537    &     1.7399       &   [-4.75, -0.76]	&	$\beta_{24}$						 					 
				& -5.5382     &             1.9633  &     [-7.91, -2.30]		   \\
				$\beta_{15}$ &   -0.7411  &         1.1475     &  [-2.77, 0.26]		&	$\beta_{25}$ 											
				&     -0.6024 &           1.0515    &     [-1.80, 0.41]		   \\
				
				$\tau_1$    &     23.7434   &        17.0624  &    [2.14, 54.96]		 &$\tau_2$  				
				&    23.7272  &       17.0331     &    [2.10, 54.86]	   	 \\   					  
				$\lambda_{1}$  &    0.0140  &     0.0111     &     [0.00, 0.03]		&	$\lambda_2$    					  
				&      0.0047   &   0.0037       &  [0.00, 0.01]	 	    \\	
				$\gamma_{1}$  &     -0.1311 &      0.1434    &  [-0.39, 0.07]		&	$\gamma_2$    					  
				&      0.0130   &     0.3364     &    [-0.50, 0.59]	  	    \\	
				$	w_{11} $ &   -0.2472   &       0.2854     &    [-0.73, 0.16]	&	$	w_{21} $  				
				&   -0.2602 &         0.4126     &   [-0.94, 0.36]	    \\				 
				$w_{12} $ &    7.8945  &      5.6776         &  [0.75, 18.27]		&$w_{22} $ 				 
				&   7.7558 &       5.5825   &   [0.69, 17.91]	  \\				 				 
				$	w_{13} $ &  1.0155     &     0.7336          &     [0.11, 2.37]		&$	w_{23} $  				 
				&   0.6779     &     0.5517         &    [-0.05, 1.69]		    \\					  
				$w_{14} $   &   1.0866     &         0.7651     &    [0.14, 2.50]		&	$w_{24} $  					  
				&    0.7978  &       0.6004       & [0.00, 1.87]		   \\			   		   
				$\xi_{11}$    &    6.1220   &           4.4319   &     [0.48, 14.19]		&	$\xi_{21}$   				   
				& -0.1610  &      0.2938   &  [-0.65, 0.29]	         \\ 
				$\xi_{12}$    &  -0.1578  &        0.2070    &  [-0.52, 0.13]	&	$\xi_{22}$     				      
				& -0.0678      &     0.2378       &  [-0.49, 0.27]	       	       \\	
				\hline 	
				\multicolumn{8}{c}{\ \ \ \ \ \ \ Same value across two classes	 }	 \\ \hline 		
				$\alpha_{11}$  & -0.2195   &           0.3365   &   [-0.79, 0.28]	 &$\alpha_{21} $  			   			   
				&   0.1438    &         0.3741      &   [-0.44, 0.77]		  \\					    
				$\alpha_{12} $&     -0.2957  &  0.3892          &     [-0.97, 0.28]	&	$\alpha_{22}$  					    
				&   -0.1694  &    0.3914        &   [-0.84, 0.39]		    \\				  				
				\hline
		\end{longtable}	}
	
\indent \quad 	Only variable $Time$  has different  trends in the   longitudinal  model between the general and time-varying  JLCMs.	Compared with the results of time-varying JLCM with optimal $K=2$ obtained using same Bayesian approach , we notice the CD4 level decreases, on average, with time for the general JLCM whereas increases, on average, with time for the comparative model.  $Gender$, $PrevOI$ and $AZT$ also have  a negative effects with time  in the longitudinal submodel  but  
$PrevOI$ accounts for a sharper decrease in the  CD4 level. We also notice similar trends except $Gender$ in terms of the effect of variables in the survival models compared to time-varying JLCM.
In particular, $PrevOI$, $AZT$ and $drug$  can increase  risks in both  proposed general and time-varying JLCM, however, $gender$  reduces risks in our general JLCM.
 Estimators of both $\bm{\alpha}_1$ and $\bm{\alpha}_2$  values in the regression submodel are not zero, which indicates that $drug$ and  $AZT$ can affect the association  in the variance-covariance matrix. Although the CIs of  $\bm{\alpha}_1$ and $\bm{\alpha}_2$ include zero values, the CI values are mostly concentrated around negative values. Perhaps this issue could be resolved with running a longer MCMC chain. Huang et al. (2011) pointed that the hypothesis $ \Sigma_{u\upsilon_{i}}=\bm{0}$ can be tested to determine whether there is a latent relationship between the longitudinal measurements and survival outcomes. According to the estimated values of $\bm{\alpha_1}$ and $\bm{\alpha_2}$, we can get the estimated variance-covariance  matrix  
	$$\begin{pmatrix}
		1.0259356 \qquad -0.5285108 \qquad -0.2562484\\
		-0.5285108 \qquad 1.2981980\qquad -0.3965044\\
		-0.2562484 \qquad-0.3965044 \qquad 1.3622013
	\end{pmatrix}$$ 
	with the correlation matrix 
	$$\begin{pmatrix}
		1.0000000 \qquad-0.4579556 \qquad-0.2167607\\
		-0.4579556  \qquad1.0000000\qquad -0.2981652\\
		-0.2167607 \qquad-0.2981652  \qquad1.0000000
	\end{pmatrix}$$ for subject 5. We can see that  $ \Sigma_{u\upsilon_{5}}=(-0.2562484, -0.3965044)^T$ and there exist negative correlations for random effects between longitudinal and survival processes. For subject 7, $$\begin{pmatrix}
		1 \qquad 0 \qquad 0\\
		0  \qquad1\qquad 0\\
		0 \qquad 0 \qquad 1
	\end{pmatrix}$$ is the correlation matrix, which implies the homogeneous random covariance matrix can be included as the special case with covariance modelling. The general JLCM can provide the assessment of the homogeneous covariance assumption.  Non-zero $\bm{\xi}_{k2}$ values indicate there is evidence of a time-varying membership probability. Both  $\bm{\xi}_{12}$ and $\bm{\xi}_{22}$ are negative, which implies the membership probability of these two classes decreases with time and class 1 gets a sharper decrease.

		{ \centerline{Table 5: 	{ Jumping behaviours  in the aids dataset based on the   general JLCM with 	$K$=2 using a Bayesian approach}
		}
	}
\setlength{\tabcolsep}{0.01mm}{
	\begin{longtable}{llllllccc}
		\hline			
		\multicolumn{3}{c}{no moves 	 }	 
		&\multicolumn{2}{c}{jumping behaviours	} 
		\\ 	\hline
		Classification     & class 1/low level   & class 2/high level\ \ \ \   & in class 1 at the final time point  & in class 2 at the final time point  
		\\ 	\hline	
		Frequency    & \  \  \ 327  \ & \ \ 68 \ \  \  \ \  \   & \ \ \ \ \ \  \ \ \ \ \ \ \ \ \ \ \  46 &  \ \ \ \ \ \ \ \ \ \ \ \ \ \ \ \ 26 
		\\
		\hline
		\multicolumn{1}{c}{		total 	 }	 
		&\multicolumn{2}{c}{ \ \ \ \ \ \ \ \ \ 395 \ \  \ \ \ \ \ \ \  \ \  \	} 
		&\multicolumn{2}{c}{ \ \ \ \ \ \ 72
		}  	\\
		\hline			
	\end{longtable} }
	\indent \quad 	Table 5 summarizes the jumping behaviors of individuals in the AIDS dataset. It is evident that the majority of patients remain stable within the same class, with only 15\% exhibiting jumping behaviors. Among those with jumping behaviors, approximately 64\% transition from the higher CD4 group (indicating better health) to the lower CD4 group, suggesting a potential decline in their health.  \\
	\indent \quad	Table 6 presents a comparison of the posterior predictive probabilities for each group between the time-varying JLCM and the proposed general JLCM under a Bayesian framework. It is observed that the majority of individuals are classified into class 1 at the final time-point for both models, with comparable percentages across the two. Both models are reasonable for the AIDS dataset, and by allowing group membership to vary over time, they effectively capture the variability in the data without requiring an additional class. However, as previously discussed, the estimation results suggest that the general JLCM provides a better fit for the data, offering stronger evidence of the effects of predictors on the relationship between the longitudinal and survival processes.\\ 

	{ \centerline{Table 6: 	{ Comparison of posterior membership probabilities between the  proposed and time-varying JLCM}
	}
}
\setlength{\tabcolsep}{3.6mm}{
\begin{longtable}{llllllccc}
	\hline			
	Model      & Estimation method&$K$& \%class1/low level &\%class2/high level  \\ 	\hline	
	time-varying JLCM (Final group) & Bayesian approach  &2  & 77.30&22.70  \\ 
	\hline	
	proposed JLCM (Final group) & Bayesian approach &2 & 79.87  &   20.13&   \\ 
	\hline		
\end{longtable}}

		\subsection{Predicted survival time to  event}	
		\indent \quad 
	Similar to the simulation study, our objective is to utilize the CD4 measurements to estimate the expected survival probabilities and assess our ability to discriminate between patients with high and low mortality risks. Using equations (11) and (12), we compute the survival predictions for eight specific subjects based on our proposed JLCM, which are presented in Figures 5 and 6. Both the longitudinal trajectories and survival probabilities are plotted together to evaluate how different jumping behaviors influence survival outcomes. The green dotted line represents the survival predictions derived from the proposed JLCM with 2 classes (the optimal model), while the blue dotted line corresponds to the survival probabilities from the time-varying JLCM with 2 classes (the best time-varying model). The vertical brown dotted and purple dashed lines denote the time of event and censoring time, respectively. In the longitudinal trajectories, black (red) circles indicate that the subject is classified in class 1 (2) at the specific time point, according to the proposed JLCM.\\
	\indent \quad Based on the survival prediction figures above, subjects 60, 263, 350, and 459, who exhibit stable trajectories with no jumps, show a gradual decline in survival probability, stabilizing around 95\% by the end of the study. For subject 60, the survival probability at the time of event, as predicted by the general JLCM (0.9933 vs 0.8367), 
	is slightly more accurate than that from the time-varying JLCM. Additionally, the general JLCM responds more rapidly to the steep decline in survival probabilities for subjects 414 and 436, adjusting accordingly. For subject 436, who remains in class 2 with a significant jump to class 1 at the final time point, the proposed JLCM predicts a lower survival probability, reflecting this transition. Similarly, subjects 188 and 455 demonstrate a jump to a higher CD4 group. Compared to the time-varying JLCM, the general JLCM delivers more accurate dynamic predictions by incorporating random effects and covariance modelling. This approach provides a more nuanced understanding of survival probabilities and highlights the relationship between the dynamic predictions of the general JLCM and the jumping behaviors observed in the longitudinal trajectories. The dynamic prediction plots for these eight subjects are presented in Figure C2 of Appendix C.\\

	\section{Conclusion and future work}
	\quad In this paper, we proposed an extended time-varying JLCM incorporating additional covariance modelling. The general JLCM introduces two additional regression submodels to describe the heterogeneous random covariance matrix, which is decomposed using the Cholesky decomposition technique. This covariance modelling allows us to assess the impact of covariates on the association between the longitudinal and survival processes, while also enabling changes in each subject's group classification over time. Results from both the simulation study and the real-data analysis of the AIDS dataset demonstrate that the general JLCM outperforms the time-varying JLCM in terms of error rates and prediction accuracy. Moreover, in dynamic survival probability predictions, the general JLCM is more responsive to changes in class membership over time compared to the time-varying JLCM, yielding more accurate survival probability forecasts.\\
	\indent \quad There are various extensions to consider for further work on the JLCM.  One limitation of the proposed models in this paper is the assumption that the distribution of the survival event only depends on the classification at the last time point.  As we have seen over both simulation and real data analysis, it might prevent our analysis from fully and accurately capturing the dynamic survival predictions. We could overcome this issue by considering the joint distribution of the longitudinal response and survival time at each time point, conditional on the longitudinal history and survival up to that point. A robust joint modelling approach may also be considered to deal with outlying longitudinal measurements. The assumption of normal distribution for the random errors can be relaxed by  a skew-normal distribution. In addition, rather than  assume that  every subgroup has the same random effects, we can get a more general and flexible model by setting class-specific random effects. For further investigation, we can also consider a more efficient method to identify the optimal $k$ value besides BIC, DIC etc. E.g. the  overfitted mixture model (Rousseau \& Mengersen, 2011) can increase the efficiency of group number selection.\\
	\subsubsection{Relax  the assumption that the distribution of the survival event only depends on the classification at the last time point.}
	\indent \quad 	In this paper, we assume that only the  classification at the last time point is taken into account for the time-to-event outcomes, which  leads to the limitation to capture the dynamic survival predictions accurately. For further improvement, we can  consider the joint distribution of the longitudinal and survival processes at each time point, assuming the longitudinal history and the fact the individual survived up to that point. Then, we can use the conditional independence assumption to separate the joint distributions. Under this assumption, we can rewrite the likelihood in   time-varying JLCM (Miao and Charalambous, 2022) for an example as below. \\
		\indent \quad  We  denote $\bm{\Psi}$ as the  collection of model parameters,   such that $\bm{\Psi}=(\bm{\xi}_{}, \bm{\beta}_{}, \bm{\omega}_{},  \bm{\delta}_{},   \bm{\tau}_{},  \bm{\lambda}_{},  \bm{\Sigma}_{ui})$ where $\bm{\xi}=(\bm{\xi}_{1}, \dots, \bm{\xi}_{K})$, $ \bm{\beta}=(\bm{\beta}_{1}, \dots, \bm{\beta}_{K})$, $\bm{\omega}=(\bm{\omega}_{1}, \dots, \bm{\omega}_{K})$, $\bm{\delta}=(\bm{\delta}_{1}, \dots, \bm{\delta}_{K})$, $\bm{\tau}=({\tau}_{1}, \dots, {\tau}_K)$, ${\lambda}=({\lambda}_{01}, \dots, \lambda_{0K})$.  The likelihood contribution for the $i$th patient is written as
	$$\begin{aligned}
		&L_i(\bm{\Psi}|\bm{R}_{i}=k, \bm{y}_{i},T_i,\ddot{\Delta}_i,\bm{U}_i)\\
		=&f(\bm{y}_{i}, T_i,\ddot{\Delta}_i|\bm{R}_{i}=k,\bm{U}_i,\bm{\Psi}) P(\bm{R}_{i}=k|\bm{\Psi}) f(\bm{U}_i|\bm{\Psi}) 
	\end{aligned} 
	$$

	Assuming that the observed data $(\bm{Y},\bm{T},\bm{\ddot{\Delta}})$ are independent conditional on the latent class indicator $R$ and the random effects $\bm{U}$, and that  every subgroup has the same random effects, considering $\bm{y}_{i}$ is observed at $t_{ij}=(t_{i1}, \dots,  t_{im_i})^T$, then we get:
$$\begin{aligned}
			&f(\bm{y}_{i}, T_i,\ddot{\Delta}_i|\bm{R}_{i}=k,\bm{U}_i,\bm{\Psi})\\
			=f&(\bm{y}_{i}|\bm{R}_{i}=k,\bm{U}_i,\bm{\Psi})f(T_i|\ddot{\Delta}_i, \bm{R}_{i}=k,\bm{U}_i,\bm{\Psi})\\
			=&\big\{\prod_{j=1}^{m_i}f(y_{ij}|R_{ij}=k, \bm{R}_{i}=k,\bm{U}_i,\bm{\Psi})\big\}f(T_i>t_{i1}|\ddot{\Delta}_i, \bm{U}_i,\bm{\Psi})\\
			&\prod_{j=1}^{m_i-1}f(T_i>t_{i(j+1)}|\ddot{\Delta}_i, R_{ij}=k,\bm{U}_i,\bm{\Psi}, T_i>t_{ij})\\
			&f(T_i|\ddot{\Delta}_i, R_{im_i}=k,\bm{U}_i,\bm{\Psi}, T_i>t_{im_i})
		\end{aligned}  
		$$
	The joint likelihood function is
	$$\begin{aligned}
		L_{}(\bm{\Psi}|\bm{R}, \bm{Y}, \bm{T},\bm{\ddot{\Delta}},\bm{U})
		=\prod_{i=1}^{N}f(\bm{U_i}|\bm{\Psi})\prod_{j=1}^{m_i}\prod_{k=1}^{K}P_{ijk} 
	\end{aligned}  
	$$
	Where
	$$\begin{aligned}
		P_{ijk}&=\left\{
		\begin{array}{ll}
			\{P(R_{ij}=k|\bm{\Psi})f(y_{i1}|R_{i1}=k, \bm{U}_i,\bm{\Psi})f(T_i>t_{i1}|\ddot{\Delta}_i, \bm{U}_i,\bm{\Psi})\\
			f(T_i>t_{i2}|\ddot{\Delta}_i, R_{i1}=k, \bm{U}_i,\bm{\Psi}, T_i>t_{i1})\}^{I(R_{i1}=k)}\\    	
			\ \ \ \ \ \ \ \ \ \ \ \ \ \ \ \ \ \ \ \ \ \ \ \ \ \ \ \ \ \ \ \ \ \ \ \ \ \ \ \ \ \ \ \ \ \ \ \ \ \ \ \ \ \ \ \ \ \ \ \ \       \text{for } j=1, t_{i1}\neq0; \\
			\{P(R_{ij}=k|\bm{\Psi})f(y_{i1}|R_{i1}=k, \bm{U}_i,\bm{\Psi})f(T_i>t_{i2}|\ddot{\Delta}_i, R_{i1}=k, \bm{U}_i,\bm{\Psi}, T_i>t_{i1})\}^{I(R_{i1}=k)}\\    	
			\ \ \ \ \ \ \ \ \ \ \ \ \ \ \ \ \ \ \ \ \ \ \ \ \ \ \ \ \ \ \ \ \ \ \ \ \ \ \ \ \ \ \ \ \ \ \ \ \ \ \ \ \ \ \ \ \ \ \ \ \       \text{for } j=1, t_{i1}=0; \\
			\{P(R_{ij}=k|\bm{\Psi})f(y_{ij}|R_{ij}=k, \bm{U}_i,\bm{\Psi})f(T_i>t_{i(j+1)}|\ddot{\Delta}_i, R_{ij}=k,\bm{U}_i,\bm{\Psi}, T_i>t_{ij})\}^{I(R_{ij}=k)}\\    	\ \ \ \ \ \ \ \ \ \ \ \ \ \ \ \ \ \ \ \ \ \ \ \ \ \ \ \ \ \ \ \ \ \ \ \ \ \ \ \ \ \ \ \ \ \ \ \ \ \ \ \ \ \ \ \ \ \ \ \ \ \text{for } 2\leq j \leq m_i-1; \\
			\{P(R_{im_i}=k|\bm{\Psi})f(y_{im_i}|R_{im_i}=k, \bm{U}_i,\bm{\Psi})f(T_i|\ddot{\Delta}_i, {R}_{im_i}=k,\bm{U}_i,\bm{\Psi}, T_i>t_{im_i})\}^{I(R_{im_i}=k)} \\    	\ \ \ \ \ \ \ \ \ \ \ \ \ \ \ \ \ \ \ \ \ \ \ \ \ \ \ \ \ \ \ \ \ \ \ \ \ \ \ \ \ \ \ \ \ \ \ \ \ \ \ \ \ \ \ \ \ \ \ \ \ \text{for }  j = m_i 
		\end{array}
		\right. 
	\end{aligned}  
	$$
	$P(R_{ij}=k)=\pi_{ijk}={\exp(\bm{X}_{1i}^T(t_{ij})\bm{\xi}_{k})}/{\sum_{s=1}^{K}\exp(\bm{X}_{1i}^T(t_{ij})\bm{\xi}_{s})}, k=1, \ldots, K$;\\
	$f(y_{ij}|(R_{ij}=k, \bm{U}_i)= (2\pi\tau_k )^{-{1}/{2}}\exp\{-{1}/({2\tau_k})(y_{ij}- \bm{X}_{2i}^T(t_{ij})\bm{\beta}_{k}-\bm{Z}_{i}^T(t_{ij})\bm{U}_i)^2\}$;\\ 
 $f(T_i|\ddot{\Delta}_i, {R}_{im_i}=k,\bm{U}_i,\bm{\Psi}, T_i>t_{im_i}) = \lambda_{ik}(t|R_{im_i}=k, \bm{U}_i,\bm{\Psi})^{\ddot{\Delta}_i} \exp(-H_{ik}(t_{}|R_{im_i}=k, \bm{U}_i,\bm{\Psi}))$; \\
	$f(\bm{U}_i|\bm{\Psi})=(2\pi)^{-{q}/{2}}det(\Sigma_{ui})^{-{1}/{2}}\exp(-{tr(\Sigma_{ui}^{-1}\bm{U}_i\bm{U}_i^T)}/{2})$.
	
	We expect that the estimation results could be more accurate with all classification information of each time point for survival submodel considered into calculation.
	
	\subsubsection{Robust longitudinal submodel }
		\indent \quad  Denote $y_{ijk}$ as a repeated measurement for subject $i$ at the $jth$ visit in the class $k$, and the longitudinal submodel for $y_{ij}$ is:
	$$\begin{aligned}
		y_{ij}|(R_{ij}=k)=\bm{X}_{2i}^T(t_{ij})\bm{\beta}_{k}+\bm{Z}_{i}^T(t_{ij})\bm{U}_{i}+\epsilon_{ijk}
	\end{aligned}  
	$$
	
	Where $\bm{X}_{2i}(t_{ij})$ is the covariate vector of subject $i$, $t_{ij}$ is denoted as the $jth$ visit time of subject $i$, time is treated as the polynomial  term in the covariate vector $\bm{X}_{2i}$; $\bm{\beta}_{k}$ is the parameter vector for class $k$; $\bm{U}_{i}$ ($q \times 1$) are the random effects for subject $i$.  $\bm{Z}_{i}(t_{ij})$ denotes the vector of covariates  at the $jth$ visit time associated with the  random effects $\bm{U}_{i}$. We assume that these two design matrices, $\bm{X}_{2i}(t_{ij})$ and $\bm{Z}_{i}(t_{ij})$,  may have time-dependent covariates measured at time points $t_i= (t_{i1}, t_{i2}, \cdots, t_{in_i})^T$.  We assume $\epsilon_{ijk}$ is independent of $\bm{U}_{i}$ (Liu et al., 2015).\\
		\indent \quad  In order to deal with some outlying data points, we propose  $\epsilon_{ijk}  \sim t(0,\tau_k^2,\nu)$ , where $\nu$ represents the degrees of freedom (Li et al., 2009). The normal distribution is included as a special case when $\nu \to\infty$ (For simplicity, we may assume $\nu$ is prespecified by the limited empirical experience in practice).  This robust longitudinal submodel can be considered as a more general longitudinal submodel than the linear mixed submodel (including both normal  and $t$  distributions).\\
		\indent \quad   Using the fact that the t-distributed error can be represented by a gamma-normal mixture distribution, we can write the  $\epsilon_{ijk}$ in the form of  $\epsilon_{ijk}=\gamma_{ijk}^{-\frac{1}{2}}\cdotp \varepsilon_{ijk}$, where $\gamma_{ijk} \sim \Gamma(\frac{\nu}{2},\frac{\nu}{2})$ and $\varepsilon_{ijk}\sim N(0,\tau_k^2)$. We assume that $\gamma_{ijk} $, $U_i$ and $\varepsilon_{ijk}$ are independent (Huang et al., 2010). This formulation can help   reduce the complexity in computation involved with the MCMC implementation for the robust joint model.
	
	\subsubsection{Selection of optimal number of classes}
		\indent \quad 	An important issue arising in joint latent class models is the selection of the optimal number of classes.  Common approaches to identify the optimal $k$ values in both Bayesian and Frequentist settings are the deviance information criterion (DIC)(Celeux et al., 2006),   Bayesian information criterion (BIC)(Schwarz et al., 1978), Bayes factor as well as reversible jump MCMC algorithm(Green, 1995) etc. The main disadvantage of these methods is the issue of recomputation for all model settings with different $k$ values. \\
	\indent \quad Apart from criterions we used in the paper (BIC, DIC etc.), Rousseau and Mengersen (2011) Criterion, which assumes an  overfitted mixture model with more latent subclasses than present in the dataset, might be implemented instead. This  criterion could be considered in our approach to avoid having to fit JLCM over several different $k$, which may make computation more efficient.\\
	 	\indent \quad 	The overfitted mixture model can  converge to the true mixture model when the non empty classes are  specified by giving a small proportion of subjects to empty classes. Firstly a large enough number of classes  is setted for the latent class submodel. Then the group number $K$ (the number of non empty classes) at every iteration $i$ can be calculted by $k_{i,opt}=K-\sum_{k=1}^{K}I(\frac{n_{ik}}{n}\leq v)$, where $k$ is the numer of class, $K$ is the total number of classifications, $n$ is the total number of subjects, $n_{ik}$ defines as the number of subjects in class $k$ at the $ith$ iteration. After getting the non empty classes at each iteration, the posterior classifications of non empty classes is calculated. This is how Rousseau and Mengersen Criterion works to refit the model with non-empty optimal number of classes.  This approach can  help  decrease the computational burden, because model need only be fitted twice (once for the highest number of classes and  once for the optimal $k$ value).

	\begin{figure}[H]
		\begin{center}
			\includegraphics[scale=0.4]{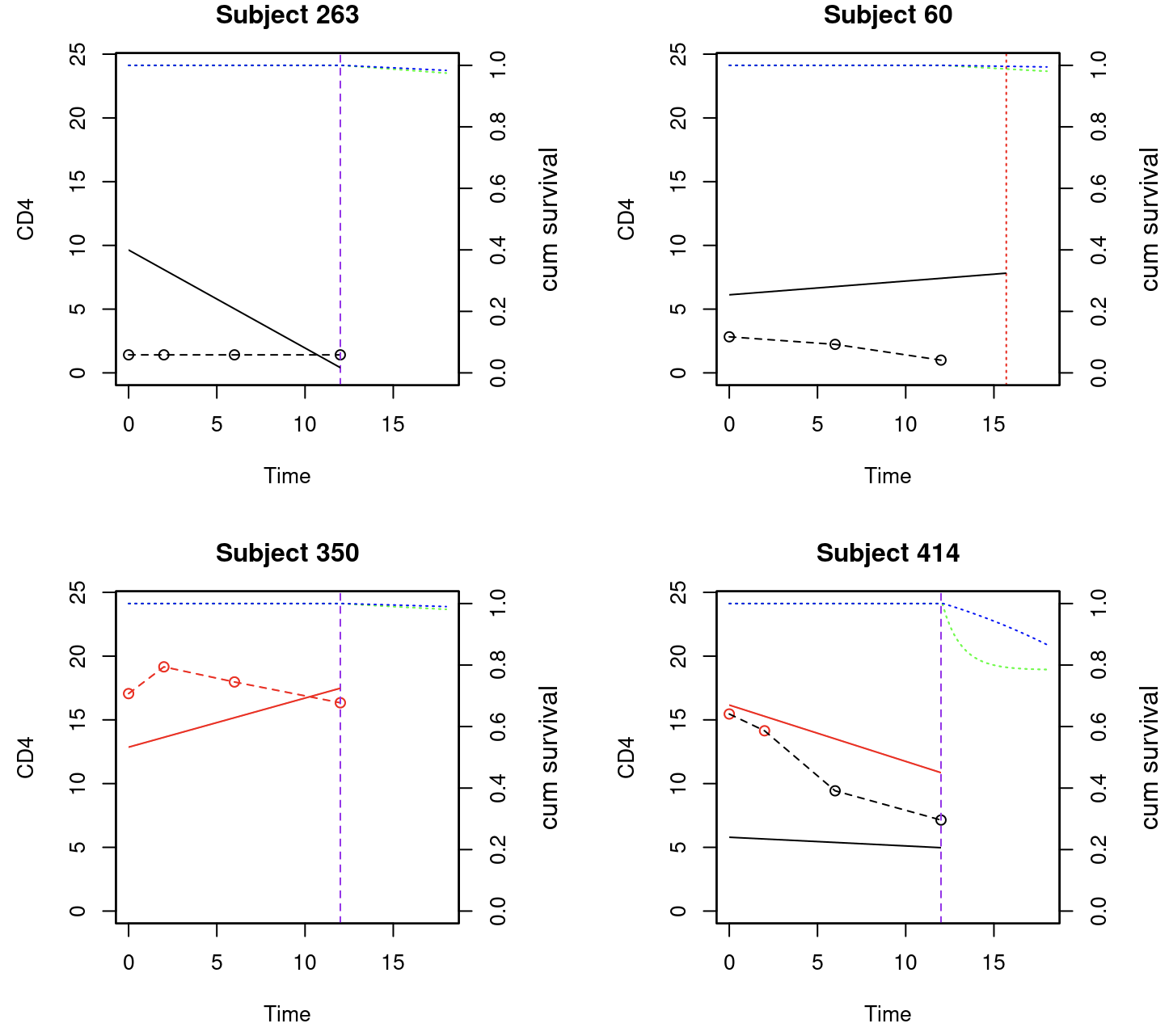}
		\end{center}
		\center{ Fig. 5 Prediction of AIDS data for four cases from both proposed and time-varying JLCM (black point: class 1/low level; red point: class 2/high level; black dashed  line: longitudinal trajectory which stayed in class 1; red  dashed  line: longitudinal trajectory which stayed in class 2; black full line: fitted trajectory which stayed in class 1; red  full line: fitted trajectory which stayed in class 2;  green dotted line: survival probability from proposed JLCM; blue dotted line: survival probability from basic JLCM; purple vertival dashed line: subject censored; red  vertical dotted line: time to event happened)
		}
	\end{figure}

	\begin{figure}[H]
		\begin{center}
			\includegraphics[scale=0.4]{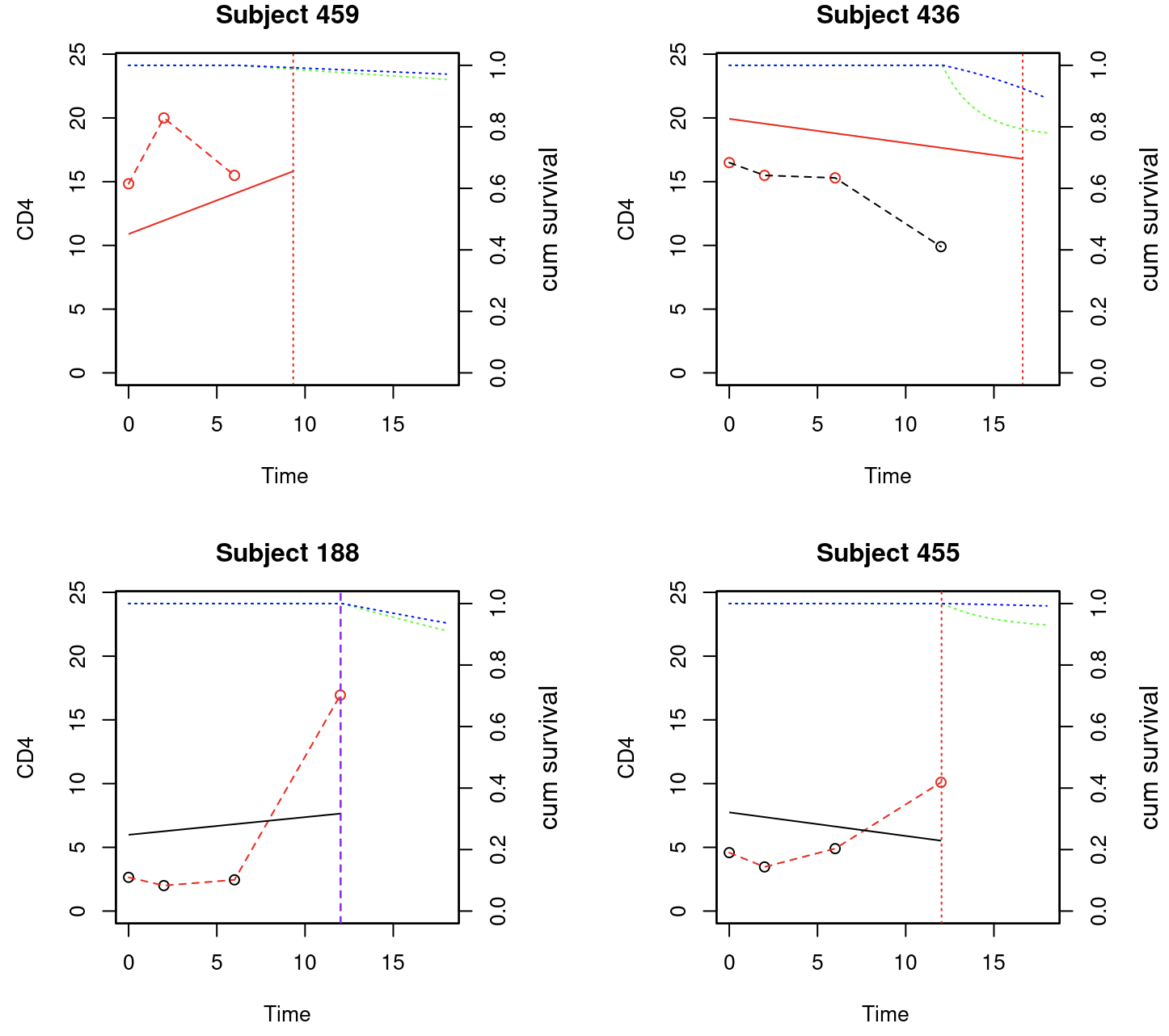}
		\end{center}
		\center{ Fig. 6 Prediction of AIDS data for four cases from both proposed and time-varying JLCM (black point: class 1/low level; red point: class 2/high level; black dashed  line: longitudinal trajectory which stayed in class 1; red  dashed  line: longitudinal trajectory which stayed in class 2; black full line: fitted trajectory which stayed in class 1; red  full line: fitted trajectory which stayed in class 2;  green dotted line: survival probability from proposed JLCM; blue dotted line: survival probability from basic JLCM; purple vertival dashed line: subject censored; red  vertical dotted line: time to event happened)
		}
	\end{figure}	
	
	\appendix 
	{\bf Appendix A: Full conditional densities}\\
	\textit{(1)Sampling $\bm{\beta}_{k}$ from normal distribution. }\\
	
		\indent \quad We set the prior distribution for $\bm{\beta}_{k}$ as $\pi(\bm{\beta}_{k}) \sim MVN(\bm{\beta}_{0}, \Sigma_{\beta_k})$ and then we can get: $\pi((\bm{\beta}_{k})	\varpropto exp(-\frac{1}{2}(\bm{\beta}_{k}-\bm{\beta}_{0})^T\Sigma_{\beta_k}^{-1}  (\bm{\beta}_{k}-\bm{\beta}_{0}))$.
	(Note: Denote $\pi(\cdot)$ as the density of the prior distribution and $\pi(\cdot|\cdot)$ as the density of the conditional distribution).\\
	
	\indent \quad  Now,  we  get the conditional distribution for $\bm{\beta}_{k}$:\\
	
	\begin{align*}
		&\pi(\bm{\beta}_{k}|\cdot)\\
		\varpropto&\prod\limits_{i=1}^N\prod\limits_{j=1}^{m_i} \Bigg\{\exp\left[-\frac{(y_{ij}- \bm{X}_{2i}^T(t_{ij})\bm{\beta}_{k}-\bm{Z}_{i}^T(t_{ij})\bm{U}_i)^2}{2\tau_k}\right]\Bigg\}^{I(R_{ij}=k)}\cdot \pi(\bm{\beta}_{k})
		\\
		\varpropto&\exp\left[-\frac{\sum\limits_{i=1}^N\sum\limits_{j=1}^{m_i}2(y_{ij}-\bm{Z}_{i}^T(t_{ij})\bm{U}_i)(\bm{X}_{2i}^T(t_{ij})\bm{\beta}_{k}){I(R_{ij}=k)}-\sum\limits_{i=1}^N\sum\limits_{j=1}^{m_i}\bm{X}_{2i}^T(t_{ij})\bm{X}_{2i}(t_{ij})\bm{\beta}_{k}^T\bm{\beta}_{k}{I(R_{ij}=k)}}{2\tau_k}\right]\\
		&\exp(-\frac{1}{2}(\bm{\beta}_{k}-\bm{\beta}_{0})^T\Sigma_{\beta_k}^{-1}  (\bm{\beta}_{k}-\bm{\beta}_{0})) \\
		=&\exp\Bigg\{-\frac{1}{2}\Bigg\{\bm{\beta}_{k}^T(\frac{\sum\limits_{i=1}^N\sum\limits_{j=1}^{m_i}\bm{X}_{2i}^T(t_{ij})\bm{X}_{2i}(t_{ij}){I(R_{ij}=k)}}{\tau_k}+\Sigma_{\beta_k}^{-1})\bm{\beta}_{k}\\
		&-2(\bm{\beta}_{k}^T\frac{\sum\limits_{i=1}^N\sum\limits_{j=1}^{m_i}(y_{ij}-\bm{Z}_{i}^T(t_{ij})\bm{U}_i)\bm{X}_{2i}^T(t_{ij}){I(R_{ij}=k)}}{\tau_k}+\Sigma_{\beta_k}^{-1}\bm{\beta}_{0})\Bigg\}\Bigg\} 
	\end{align*}

\indent \quad If $\sum_{i=1}^N\sum_{j=1}^{m_i}I(R_{ij}=k)=0$, which means the likelihood makes no contributions to the posterior distribution, we sample $\bm{\beta}_{k}$ from the prior distribution $\pi(\bm{\beta}_{k})$. \\
	
	\indent \quad We know that if $y$ has the pdf $g(y)  \varpropto exp\{-{1}/{2}(Ay^2-2By)\}$, then we can say $y \sim N({B}/{A}, {1}/{A})$. And, for $\bm{\beta}_{k}$,  if $\sum_{i=1}^N\sum_{j=1}^{m_i}I(R_{ij}=k)\neq0$,
	$$\bm{\beta}_k|\cdot \sim N\left(\frac{B}{A}, \frac{1}{A}\right)$$
	Where $ A={\sum_{i=1}^N\sum_{j=1}^{m_i}\bm{X}_{2i}^T(t_{ij})\bm{X}_{2i}(t_{ij})I(R_{ij}=k)}/{\tau_k}+\Sigma_{\beta_k}^{-1}, B={\sum_{i=1}^N\sum_{j=1}^{m_i}(y_{ij}-\bm{Z}_{i}^T(t_{ij})\bm{U}_i)\bm{X}_{2i}^T(t_{ij})I(R_{ij}=k)}/{\tau_k}+\Sigma_{\beta_k}^{-1}\bm{\beta}_{0}$.\\
	
		\textit{(2)Sample $\lambda_{0k}$ from  Gamma distribution.}\\
		
	\indent \quad Specify the inverse Gamma prior for the gamma prior for  $\lambda_{0k}$, leading to the conjugate posterior for $\lambda_{0k}$. i.e. $\pi(\lambda_{0k})$ follows Gamma distribution.  And then we can get $\pi(\lambda_{0k}) \sim \Gamma(\alpha_{\lambda},\beta_{\lambda})$.	We can get the posterior distribution of $\lambda_{0k}$ as:\\
	
		\begin{align*}
		&\pi(\lambda_{0k}^{}|\cdot)\\
		\varpropto &\Bigg\{\prod_{i=1}^{N}{\lambda_{0k}^{}}^{I(T_i^*\le C_i)I(R_{ij}=k)}
		\exp\Bigg\{-I(R_{ij}=k)\lambda_{0k}^{}\int_{0}^{t_{}}
	\exp(\gamma_k(u))\exp(\bm{X}_{3i}^T(u)\bm{\omega}_{k}+\bm{W}_i[q+1])du\Bigg\}\Bigg\} \pi(\lambda_{0k})\\
		=&{\lambda_{0k}^{}}^{\sum_{i=1}^{N}I(T_i^*\le C_i, R_{ij}=k)}
		\exp\Bigg\{-\lambda_{0k}^{}\sum_{i=1}^{N}I(R_{ij}=k)\int_{0}^{t_{}}
	\exp(\gamma_k(u))\exp(\bm{X}_{3i}^T(u)\bm{\omega}_{k}+\bm{W}_i[q+1])du\Bigg\}(\lambda_{0k}^{})^{\alpha_{\lambda}-1}\exp(-{\beta_{\lambda}}{\lambda_{0k}^{}})\\
		=&{\lambda_{0k}^{}}^{\sum_{i=1}^{N}I(T_i^*\le C_i, R_{ij}=k)+\alpha_{\lambda}-1}
	\exp\Bigg\{-\lambda_{0k}^{}(\sum_{i=1}^{N}I(R_{ij}=k)\int_{0}^{t_{}}
	\exp(\gamma_k(u))\exp(\bm{X}_{3i}^T(u)\bm{\omega}_{k}+\bm{W}_i[q+1])du+{\beta_{\lambda}})\Bigg\}
	\end{align*}
	So we can see that if $\sum I(R_{ij}=k)\neq0$,\\
	$$\lambda_{0k}^{}|\cdot \sim \Gamma(E,F)$$
	where $$E=\alpha_{\lambda}+\sum_{i=1}^{N}I(T_i^*\le C_i, R_{ij}=k)+1$$ $$F=\sum_{i=1}^{N}I(R_{ij}=k)\int_{0}^{t_{}}
	\exp(\gamma_k(u))\exp(\bm{X}_{3i}^T(u)\bm{\omega}_{k}+\bm{W}_i[q+1])du+\beta_{\lambda}$$
	$$	\bm{W}_i=(	\bm{U}_{i}, \upsilon_{i})^T	$$	
	
	If $\sum I(R_{ij}=k)=0$, $\lambda_{0k}^{}|\cdot \sim \pi(\lambda_{0k})$.\\

		\textit{(3)Sampling $\tau_k$ from inverse gamma distribution.}\\
		
	\indent \quad Further specify the inverse Gamma prior for the variance of the measurement error $\tau_k$, 
	i.e. we set the prior distribution for $\tau_k$: $\pi(\tau_k) \sim InverseGamma(\alpha^*,\beta^*) $ and then we can write: $\pi(\tau_k)=\frac{{\beta^*}^{\alpha^*}}{\Gamma(\alpha^*)}(\tau_k)^{-\alpha^*-1}\exp(-\frac{\beta^*}{\tau_k})$ ($\alpha^*>0, \beta^*>0$).
	Thus, we can get the conditional distribution for $\tau_k$:\\
	
	\begin{align*}
		&\pi(\tau_k|\cdot)\\
		\varpropto&\Bigg\{\prod_{i=1}^{N}\prod_{j=1}^{m_i}\left[(\tau_k)^{(-\frac{1}{2})}\exp\left[-\frac{(y_{ij}- \bm{X}_{2i}^T(t_{ij})\bm{\beta}_{k}-\bm{Z}_{i}^T(t_{ij})U_i)^2}{2\tau_k}\right]\right]^{I(R_{ij}=k)}\Bigg\}
		\cdot(\tau_k)^{-\alpha^*-1} \exp(-\frac{\beta^*}{\tau_k})\\
		=&\Bigg\{\prod_{i=1}^{N}\prod_{j=1}^{m_i}(\tau_k)^{-\frac{I(R_{ij}=k)}{2}}\Bigg\}(\tau_k)^{-\alpha^*-1}\exp\Bigg\{-\frac{\frac{\sum\limits_{i=1}^N\sum\limits_{j=1}^{m_i}(y_{ij}- \bm{X}_{2i}^T(t_{ij})\bm{\beta}_{k}-\bm{Z}_{i}^T(t_{ij})\bm{U}_i)^2}{2}I(R_{ij}=k)+\beta^*}{\tau_k}\Bigg\}\\
		=&(\tau_k)^{-\frac{\sum_{i=1}^{N}\sum_{j=1}^{m_i}I(R_{ij}=k)}{2}-\alpha^*-1}\exp\Bigg\{-\frac{\frac{\sum\limits_{i=1}^N\sum\limits_{j=1}^{m_i}(y_{ij}- \bm{X}_{2i}^T(t_{ij})\bm{\beta}_{k}-\bm{Z}_{i}^T(t_{ij})\bm{U}_i)^2I(R_{ij}=k)}{2}+\beta^*}{\tau_k}\Bigg\}
	\end{align*}
	
	So \begin{align*}
		\tau_k|\cdot \sim InverseGamma(&\frac{\sum_{i=1}^{N}\sum_{j=1}^{m_i}I(R_{ij}=k)}{2}+\alpha^*,\\
		&\frac{\sum\limits_{i=1}^N\sum\limits_{j=1}^{m_i}(y_{ij}- \bm{X}_{2i}^T(t_{ij})\bm{\beta}_{k}-\bm{Z}_{i}^T(t_{ij})\bm{U}_i)^2I(R_{ij}=k)}{2}+\beta^*)
	\end{align*}
	\quad If $\sum\limits_{i=1}^N\sum\limits_{j=1}^{m_i}I(R_{ij}=k)=0$, we sample $\tau_k$ from $InverseGamma(\alpha^*,\beta^*)$; if not, we sample $\tau_k$ from\\ $InverseGamma(\frac{\sum_{i=1}^{N}\sum_{j=1}^{m_i}I(R_{ij}=k)}{2}+\alpha^*,
	\frac{\sum\limits_{i=1}^N\sum\limits_{j=1}^{m_i}(y_{ij}- \bm{X}_{2i}^T(t_{ij})\bm{\beta}_{k}-\bm{Z}_{i}^T(t_{ij})\bm{U}_i)^2I(R_{ij}=k)}{2}+\beta^*)$.\\

	{\bf Appendix B:  Estimation results for real data application on  time-varying JLCM}\\

\indent \quad We use time-varying JLCM with bayesian  estimation  method to do the inference. We set time-varying  JLCM with group numbers from 1 to 3.   Table 3 in real data application  shows the summary information for time-varying JLCMs of AIDS data with various k settings. AIDS data with 2 classes  gets the smallest DIC value among three model settings. Time-varying JLCM with two subgroups is  picked up as the  optimal model here. Estimation results of time-varying JLCM with $K=2$ are shown in Table B1.\\

	{ \centerline	{Table B1:	{Estimation results for the analysis of the AIDS data based on  time-varying JLCM with $K$=2
			}
		}
	}	
\setlength{\tabcolsep}{3.3mm}{
	\begin{longtable}{llllllcc}
		\hline
		\multicolumn{4}{c}{Class 1}	 
		&\multicolumn{4}{c}{Class 2} 
		\\ 
		\hline
		Parameter           & Estimate  &sd  & CI(89\% ETI)& 	Parameter  &Estimate &sd  & CI(89\% ETI)\\ \hline 
		$\beta_{11}$ &   9.6901    &       1.5798      &   [6.61, 11.60]	
		&	$\beta_{21}$ &10.2169 &2.15924&  [6.90, 13.52]	\\
		$\beta_{12}$   &  0.0690    &         0.2261 &    [-0.18, 0.50]	
		&	$\beta_{22}$   
		& 0.0656&   0.2329&  [-0.21, 0.50]   	 \\
		$\beta_{13}$&    -0.2238  &          1.2406      &   [-2.00, 1.92]		&$\beta_{23}$  				      
		&   -0.2663 &     1.3244     &  [-2.14, 1.93]     \\
		$\beta_{14}$ &  -3.2468    &    1.7205     &   [-5.28, -0.03]	&	$\beta_{24}$						 					 
		&  -3.5140  &             2.2258  &    [-6.73, 0.13]	   \\
		$\beta_{15}$ &   -0.8531  &         1.3973   &  [-3.46, 0.81]		&	$\beta_{25}$ 											
		&    -0.9335&         1.7431  &   [-3.70, 0.77]	   \\
		
		$\tau_1$    &     42.1118   &        35.2914  &  [4.16, 112.67]	 &$\tau_2$  				
		&    43.0151 &      36.1066    &  [4.43, 115.30]  	 \\   					  
		$\lambda_{1}$  &     0.1192  &    0.0866 &    [0.01, 0.28]	&	$\lambda_2$    					  
		&     0.6856  &   0.5287   & [0.07, 1.69]	    \\	
		$\gamma_{1}$  &     4.9863 &     8.0815  & [-5.12, 19.47]	&	$\gamma_2$    					  
		&      -0.8465 &     1.5649  &    [-3.66, 1.13]	  	    \\	
		$	w_{11} $ &   0.7213  &      1.7076 &  [-1.58, 3.75]	&	$	w_{21} $  				
		& -1.2827&         1.5676    & [-4.20, 0.55]    \\				 
		$w_{12} $ &     14.7123 &     12.1755        &   [1.59, 39.05]	&$w_{22} $ 				 
		&    14.7653&      12.1698  &   [1.64, 39.24] \\				 				 
		$	w_{13} $ &   3.0325   &    3.6380      &    [-1.27, 9.90]	&$	w_{23} $  				 
		&   0.8254   &      1.3862       &   [-1.25, 3.20]		    \\					  
		$w_{14} $   &  1.1191  &      1.2803  &    [-0.45, 3.44]	&	$w_{24} $  					  
		&   1.4222&       1.2401    & [0.08, 3.84]	   \\			   		   
		$\xi_{11}$    &    12.0947  &          10.0225  &   [1.42, 32.15]	&	$\xi_{21}$   				   
		& 0.2629  &       0.7500  &  [-0.78, 1.58]         \\ 
		$\xi_{12}$    &  1.2660  &        1.5814  &  [-0.63, 4.23]&	$\xi_{22}$     				      
		& 0.4070   &     0.5235   &  [-0.24, 1.39]     	       \\	
		\hline 	
\end{longtable}	}

		{\bf Appendix C: Dynamic prediction plots of simulation and real data application}\\
		
	\begin{figure}[H]
		\begin{center}
			\includegraphics[scale=0.4]{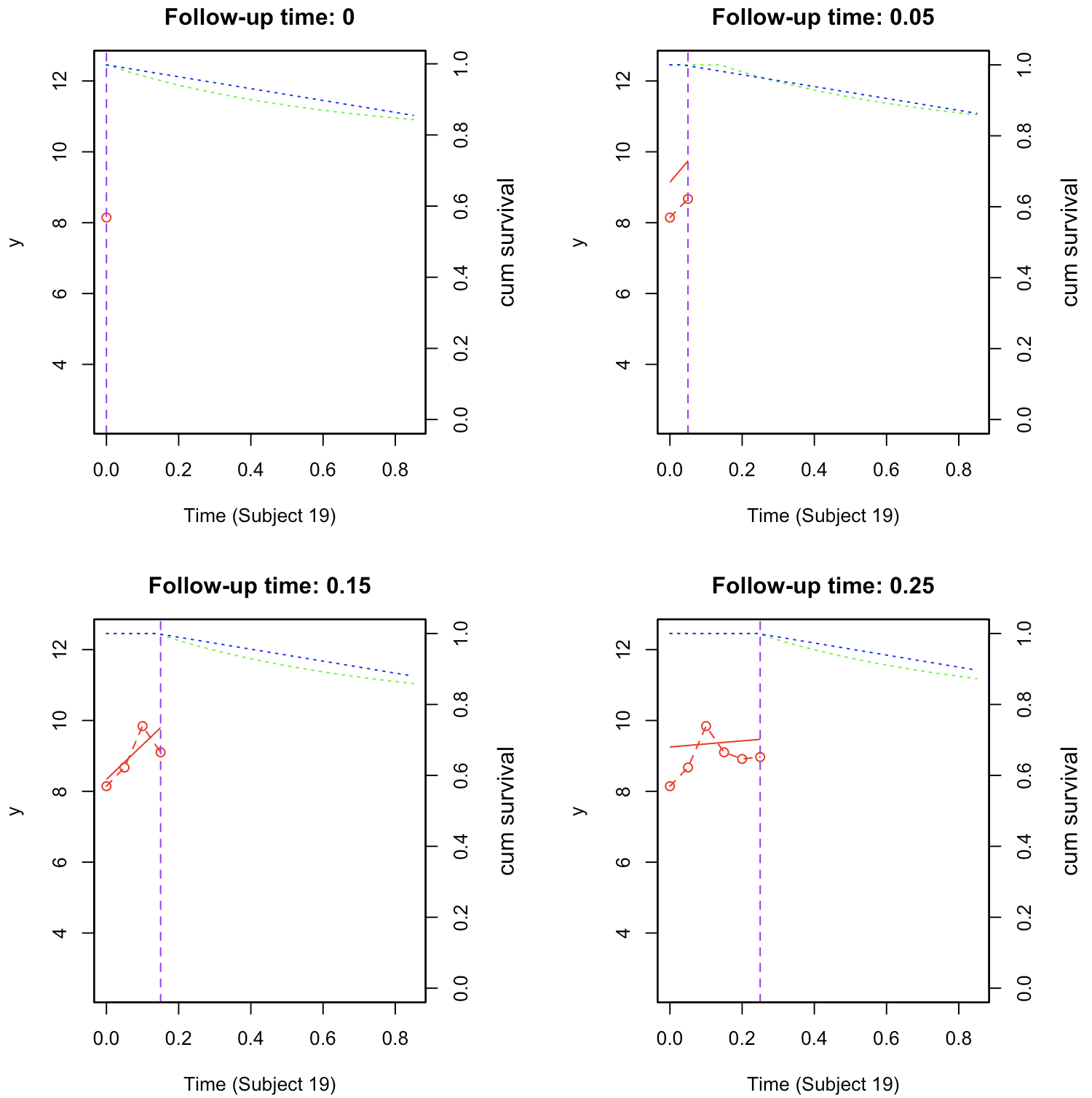}
		\end{center}
	\end{figure}
	\begin{figure}[H]
		\begin{center}
			\includegraphics[scale=0.4]{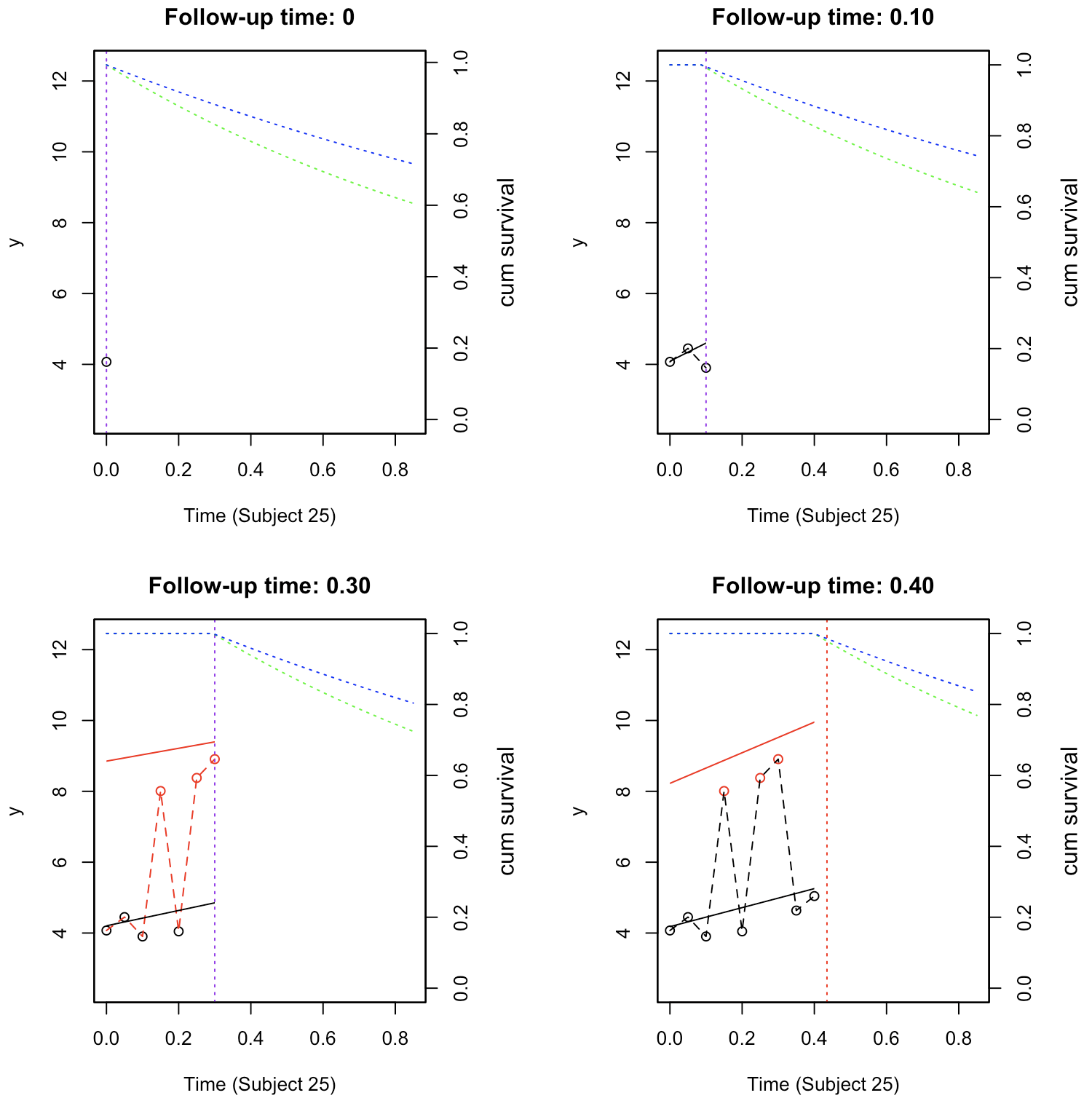}
		\end{center}
	\end{figure}
	\begin{figure}[H]
		\begin{center}
			\includegraphics[scale=0.4]{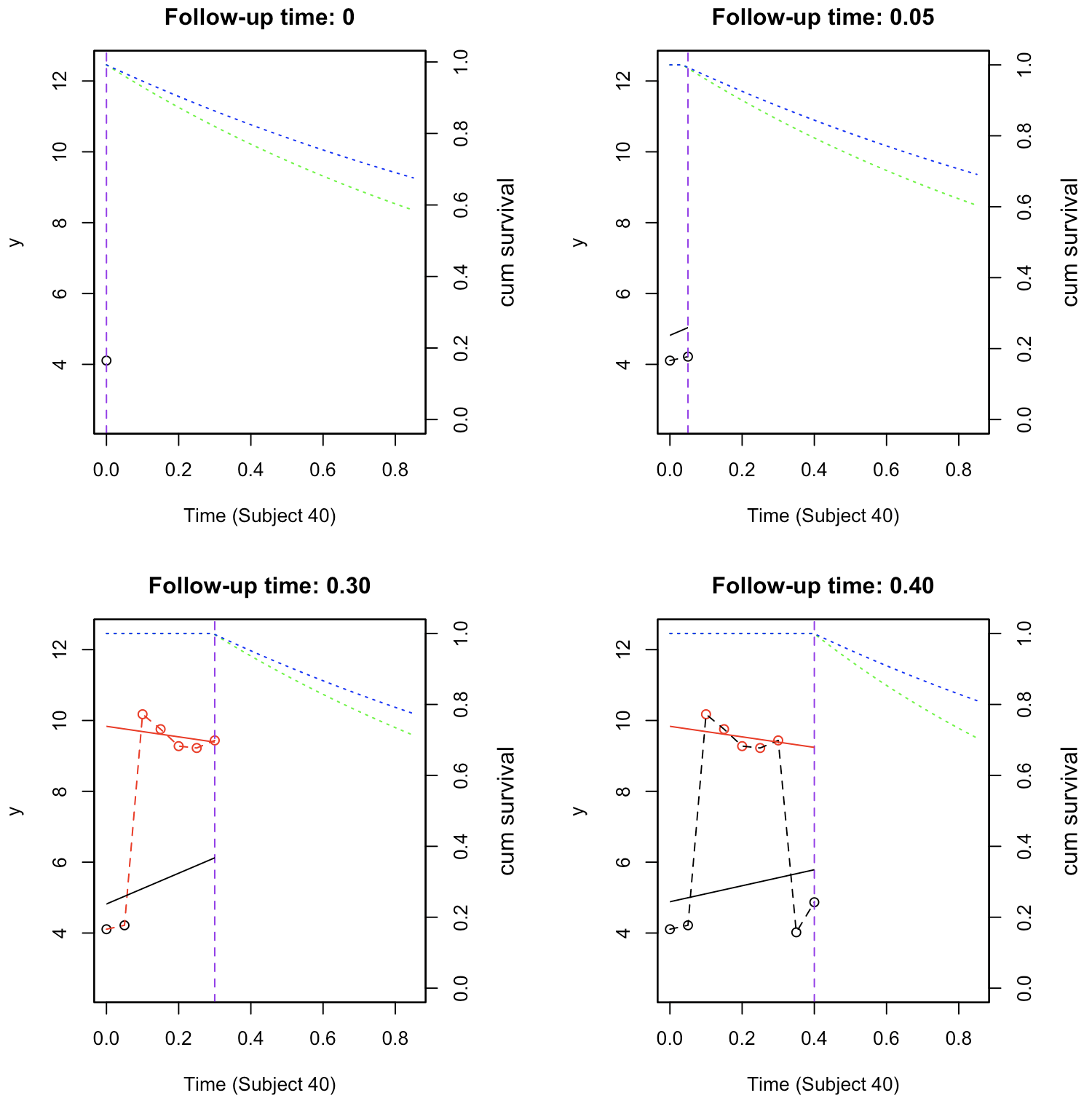}
		\end{center}
	\end{figure}
	\begin{figure}[H]
		\begin{center}
			\includegraphics[scale=0.4]{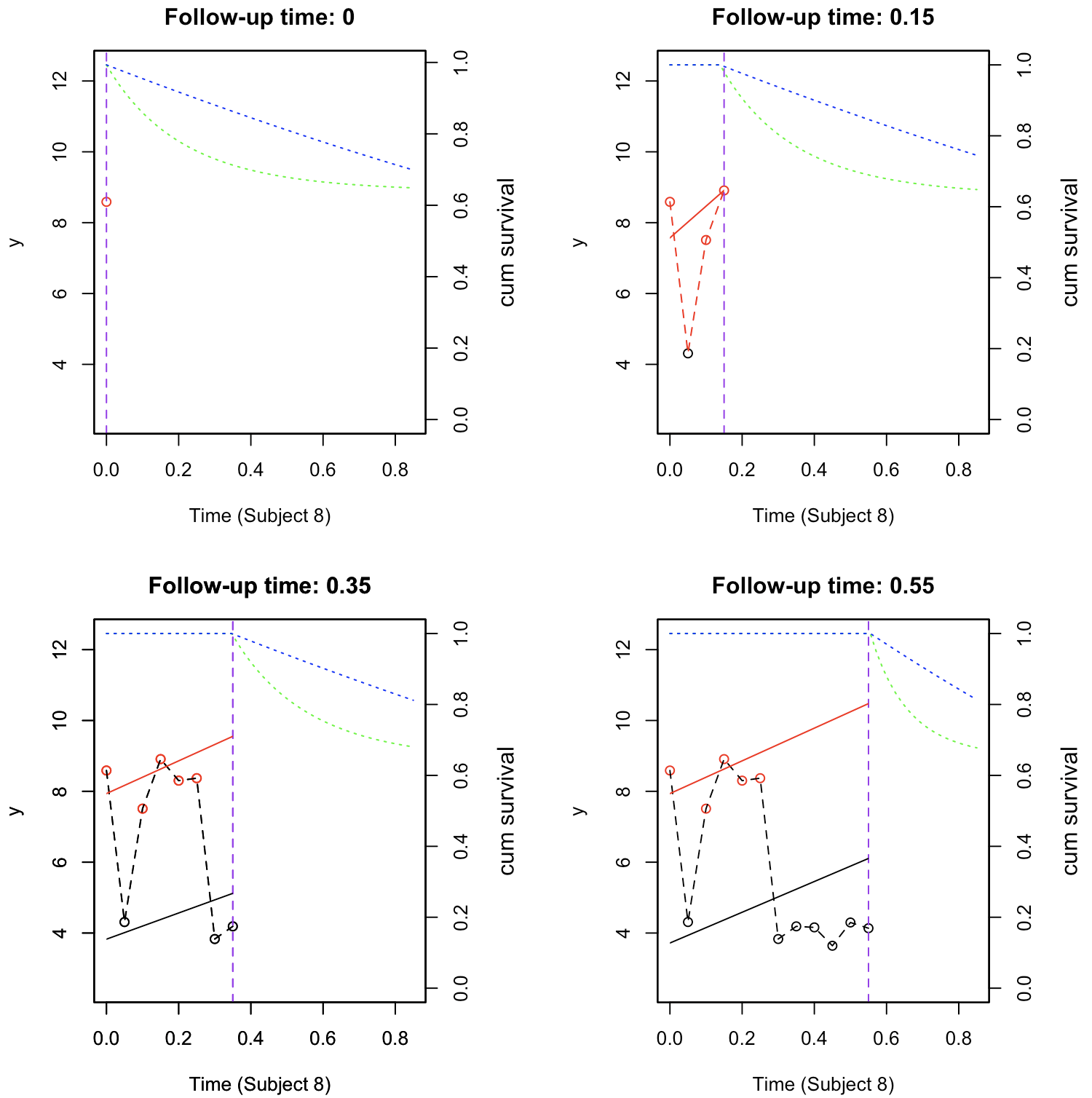}
		\end{center}
	\end{figure}
	\begin{figure}[H]
		\begin{center}
			\includegraphics[scale=0.38]{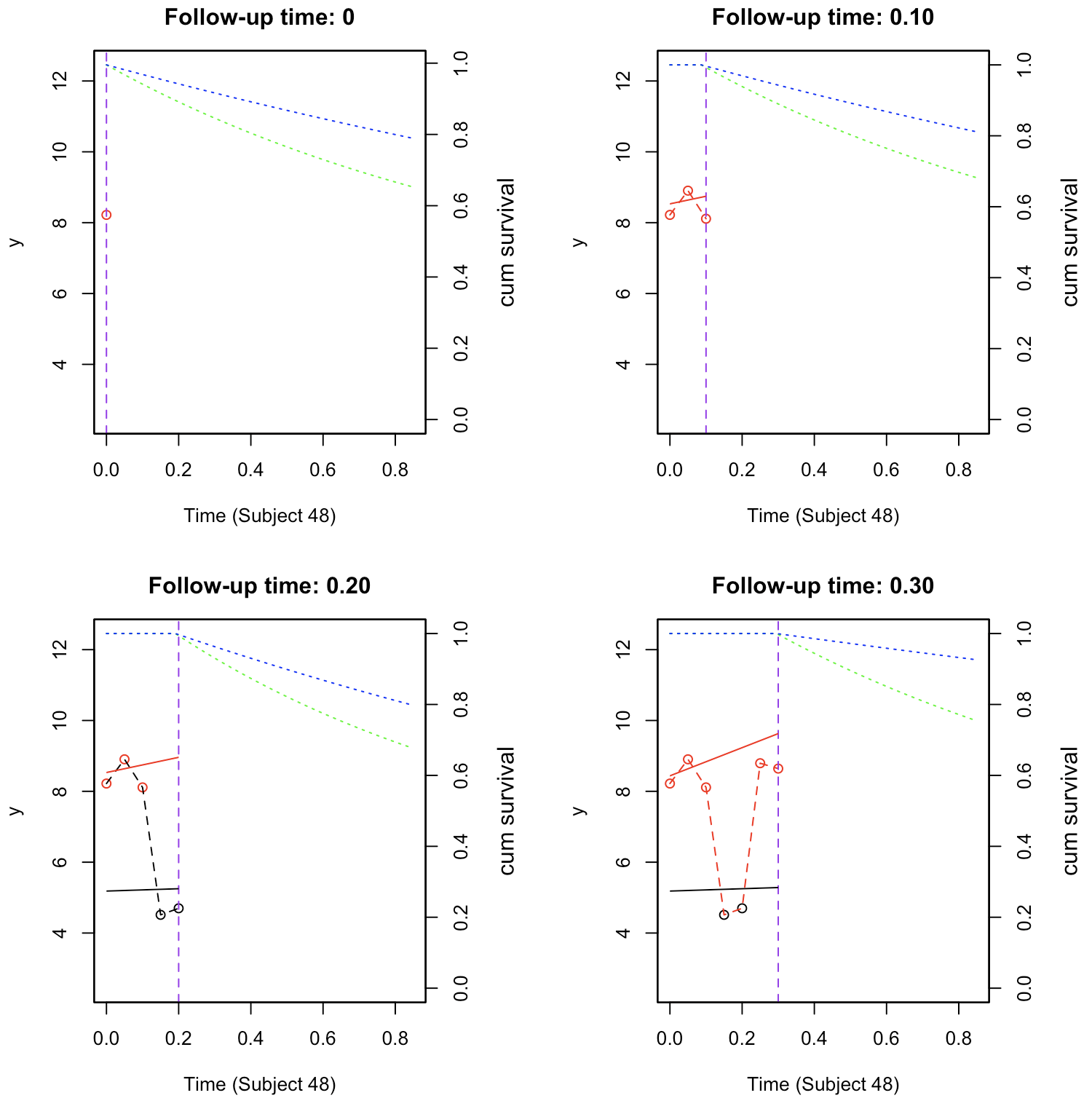}
		\end{center}
	\end{figure}
	\begin{figure}[H]
		\begin{center}
			\includegraphics[scale=0.38]{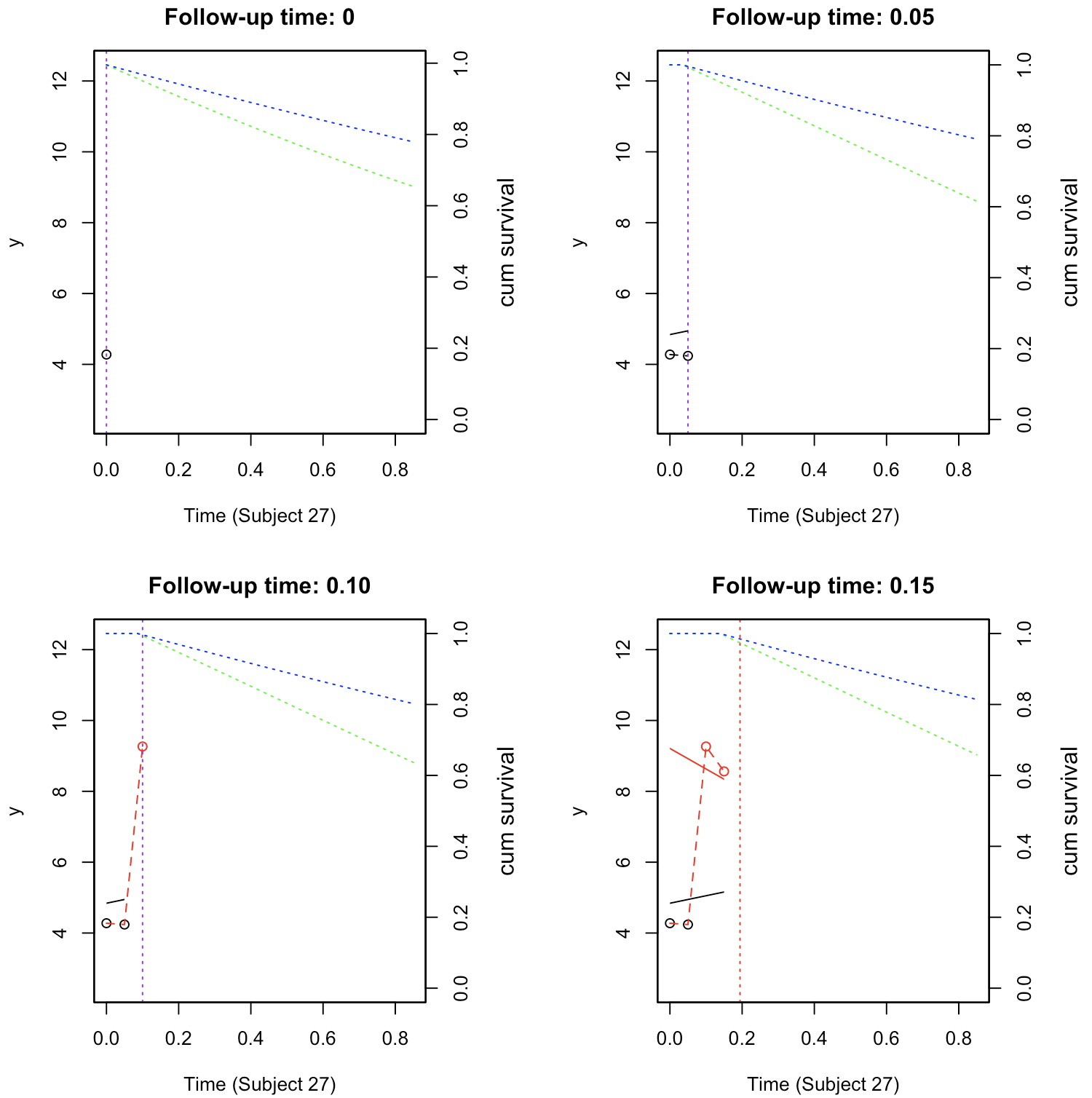}
			\center{ Fig. C1. Dynamic Survival predictions of simulated data for six cases from both proposed and basic JLCM (black point: class 1/low level; red point: class 2/high level; black dashed  line: longitudinal trajectory which stayed in class 1; red  dashed  line: longitudinal trajectory which stayed in class 2; black full line: fitted trajectory which stayed in class 1; red  full line: fitted trajectory which stayed in class 2;  green dotted line: survival probability from proposed JLCM; blue dotted line: survival probability from basic JLCM; purple vertival dashed line: subject censored; red  vertical dotted line: time to event happened)  }
		\end{center}
	\end{figure}

	\begin{figure}[H]
		\begin{center}
			\includegraphics[scale=0.4]{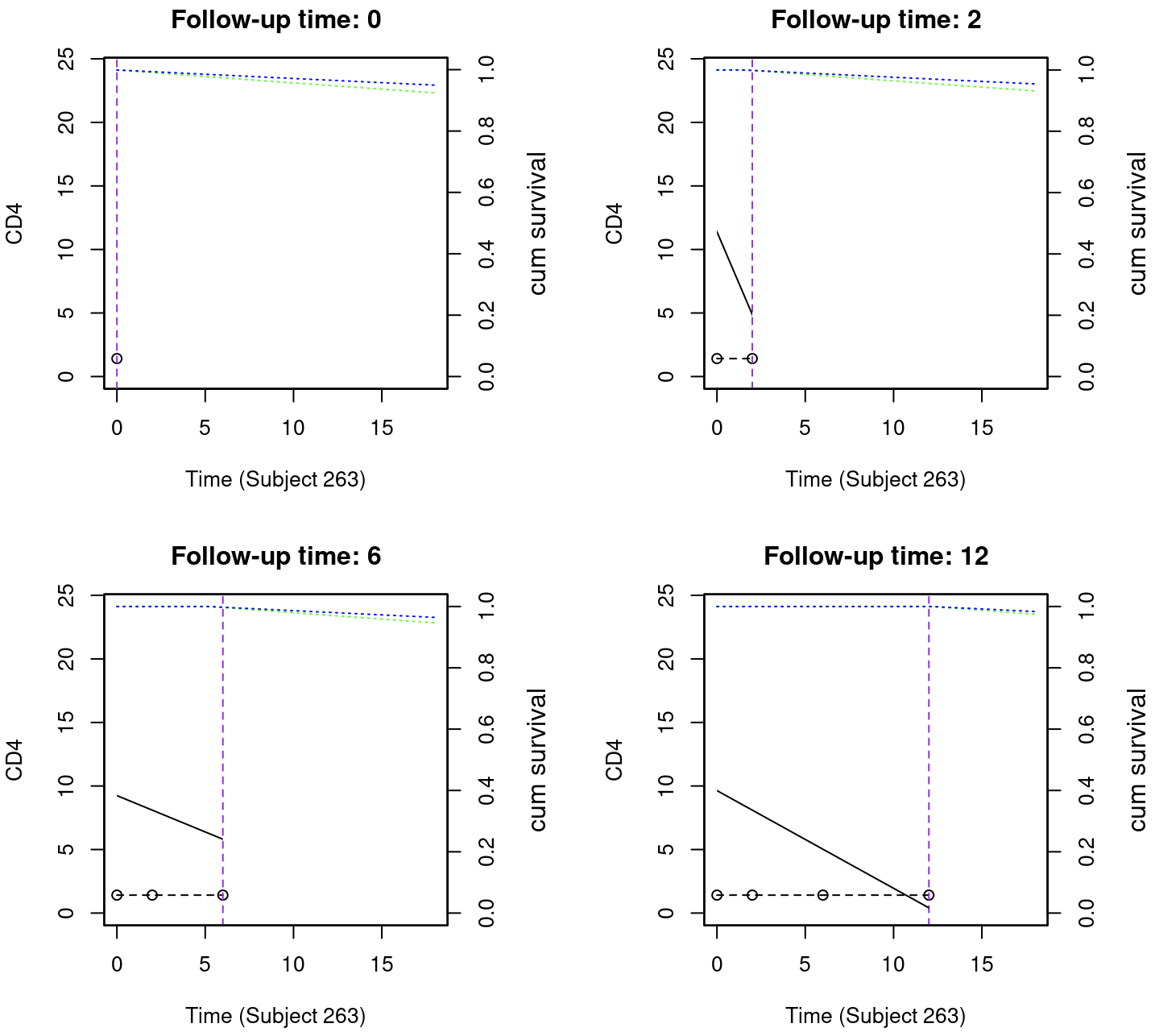}
		\end{center}
	\end{figure}
	
	\begin{figure}[H]
		\begin{center}
			\includegraphics[scale=0.4]{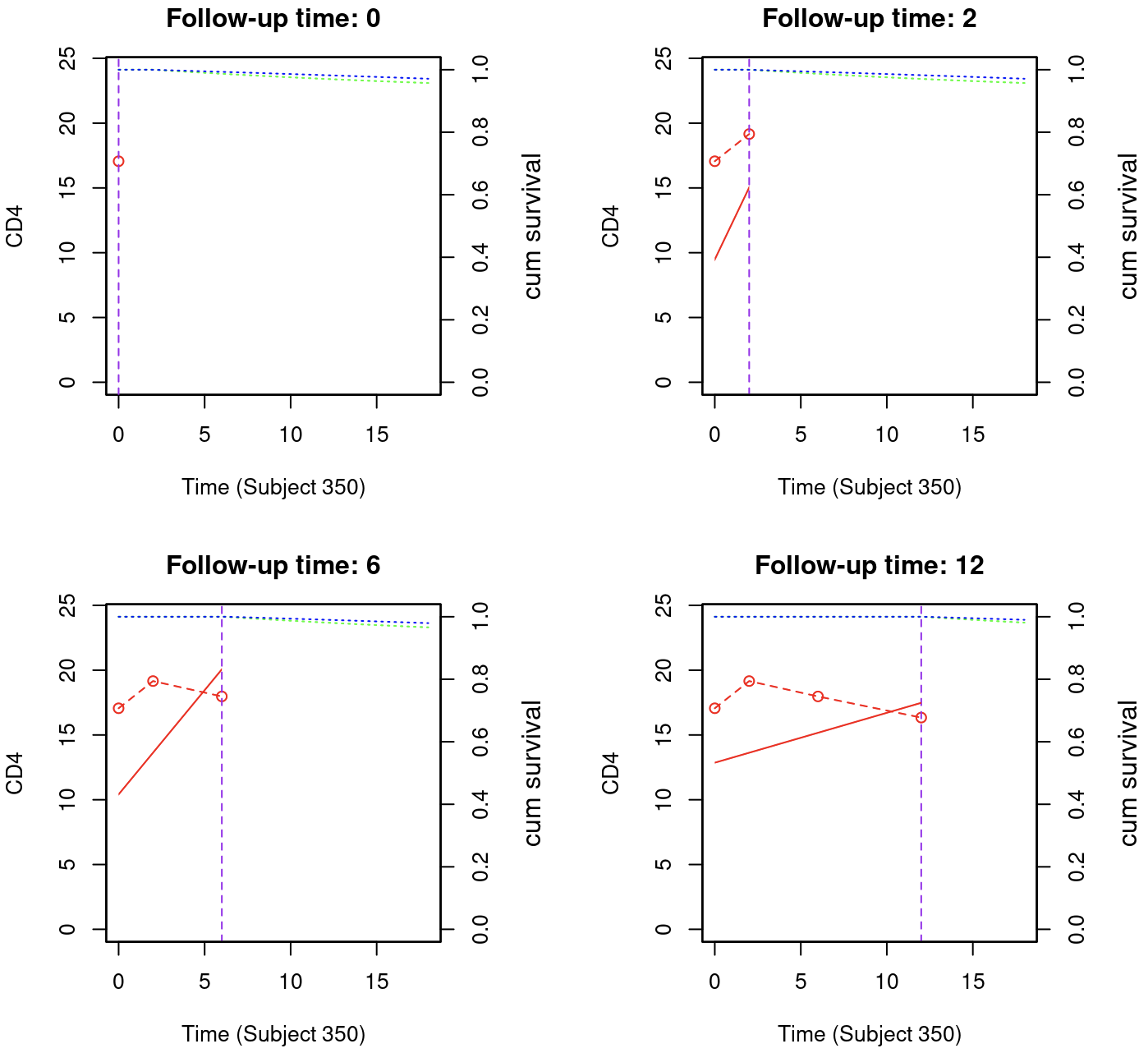}
		\end{center}
	\end{figure}

	\begin{figure}[H]
		\begin{center}
			\includegraphics[scale=0.35]{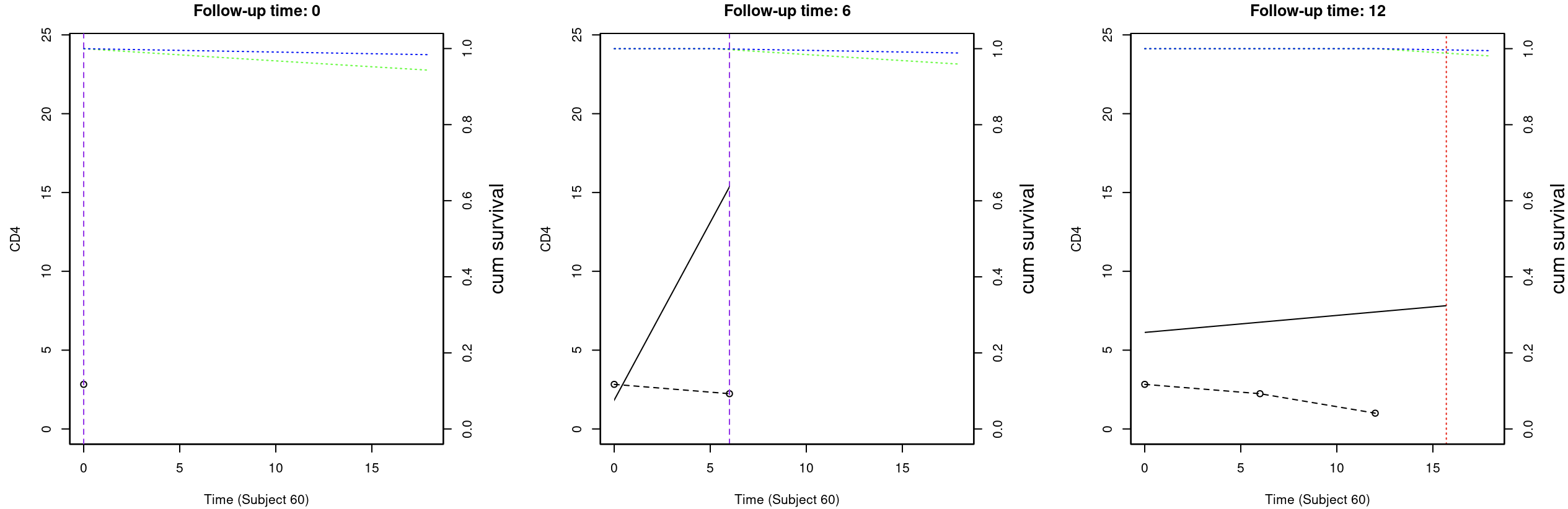}
		\end{center}
	\end{figure}
	
	\begin{figure}[H]
		\begin{center}
			\includegraphics[scale=0.42]{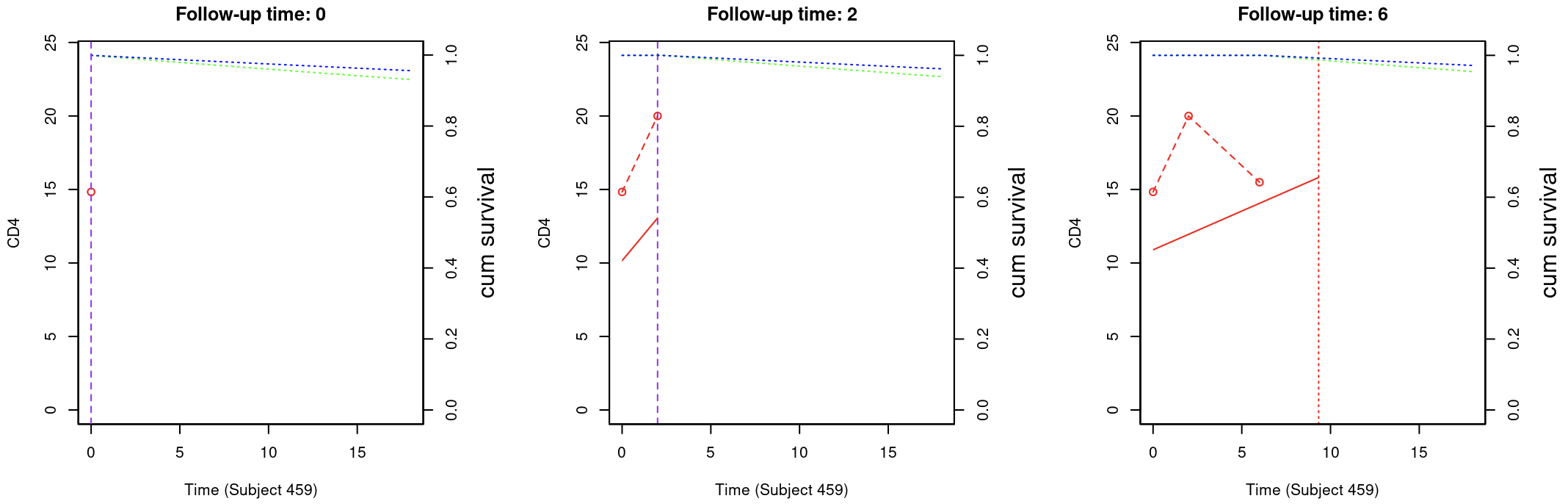}
		\end{center}
	\end{figure}

	\begin{figure}[H]
		\begin{center}
			\includegraphics[scale=0.4]{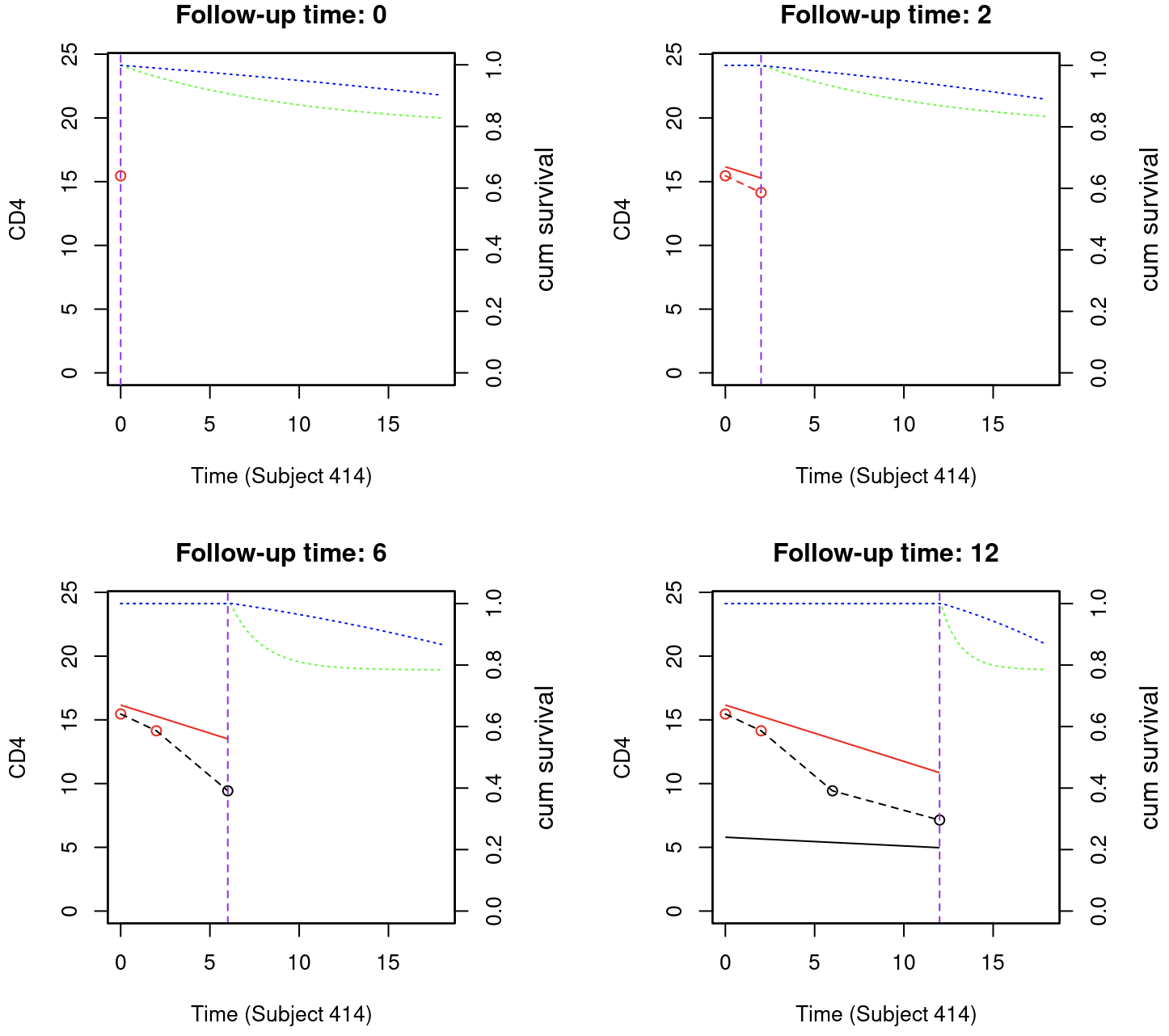}
		\end{center}
	\end{figure}

	\begin{figure}[H]
		\begin{center}
			\includegraphics[scale=0.4]{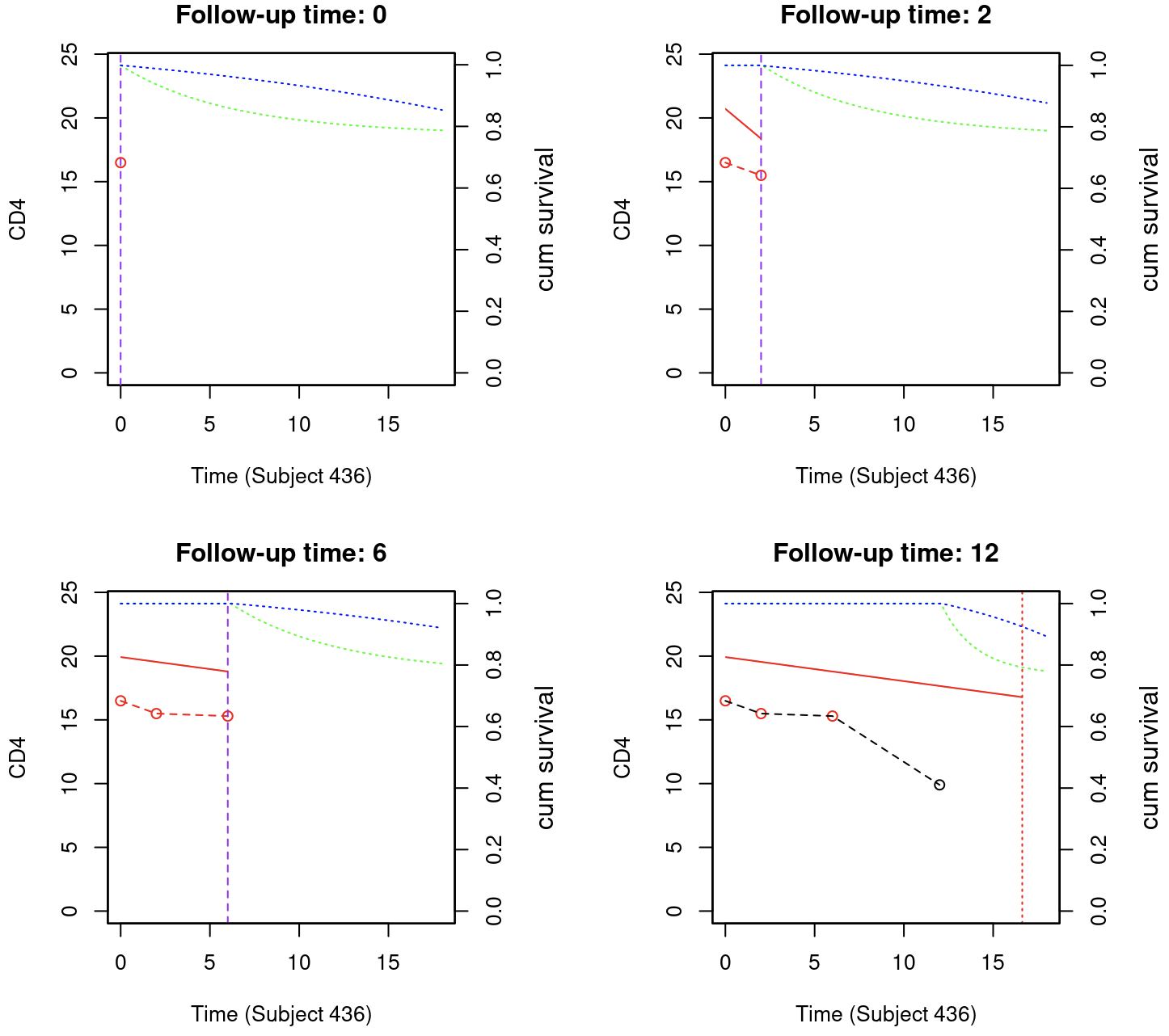}
		\end{center}
	\end{figure}

	\begin{figure}[H]
		\begin{center}
			\includegraphics[scale=0.4]{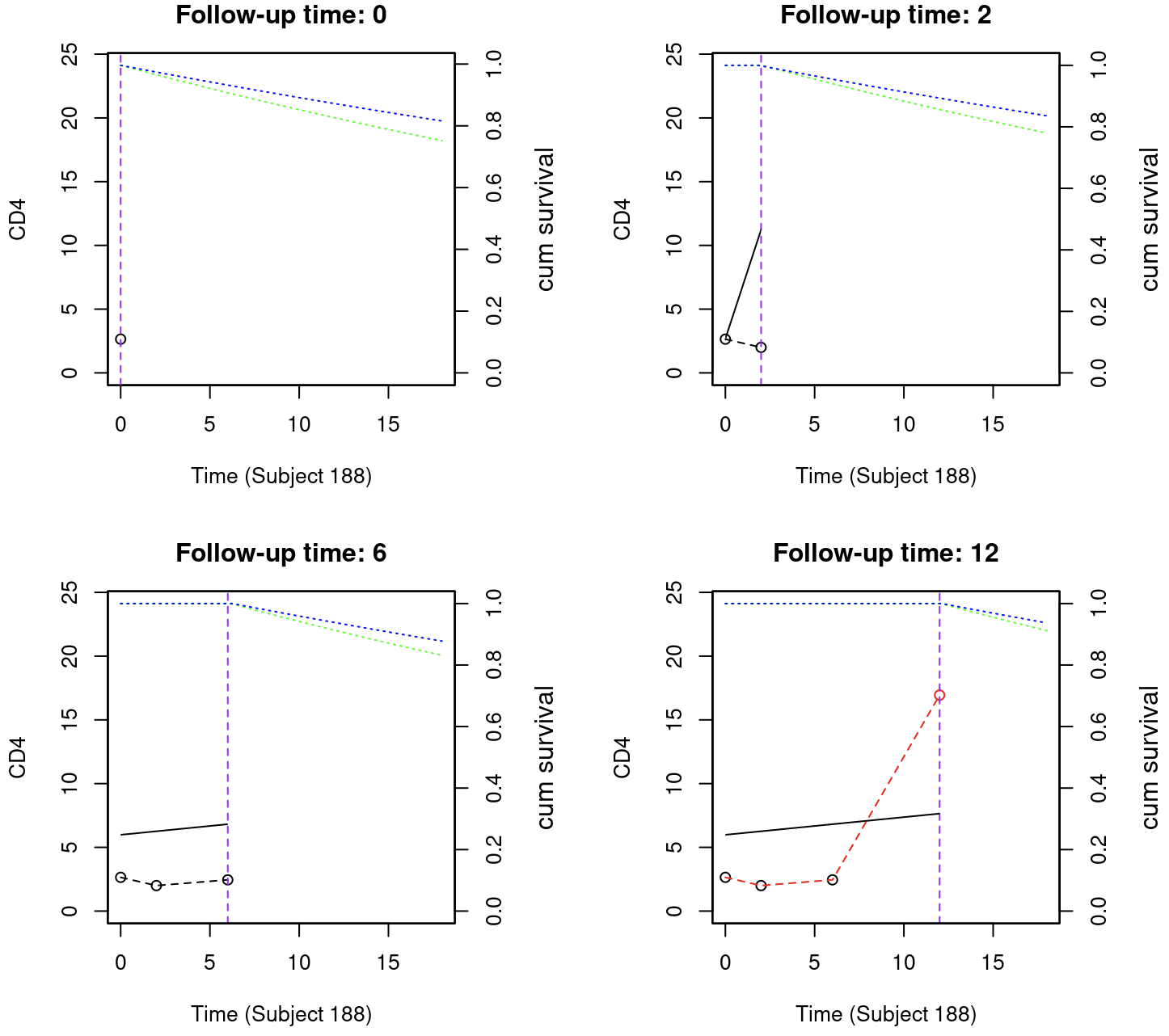}
		\end{center}
	\end{figure}

	\begin{figure}[H]
		\begin{center}
			\includegraphics[scale=0.4]{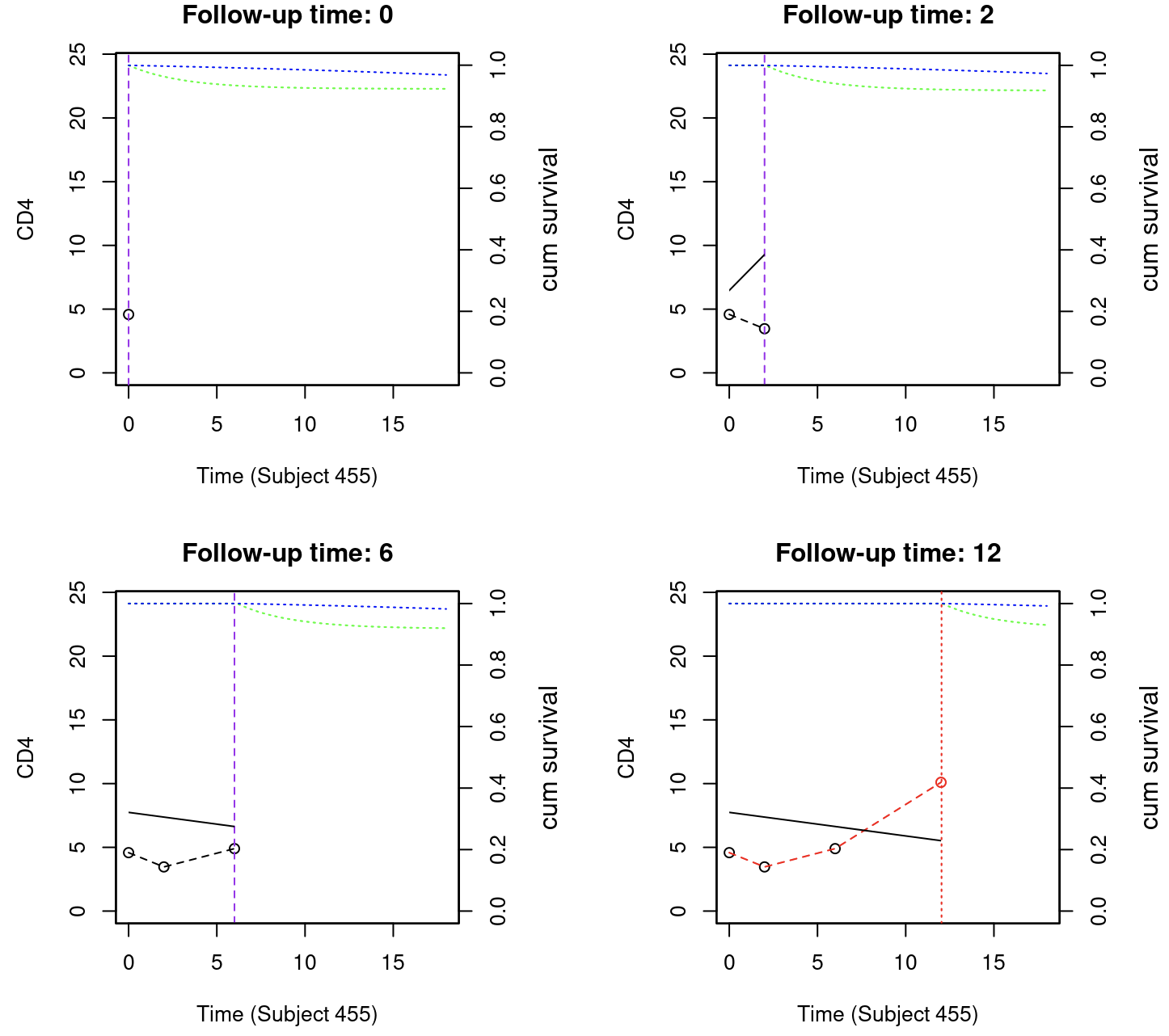}
			\center{ Fig. C2. Dynamic Survival predictions of simulated data for eight cases from both proposed and basic JLCM (black point: class 1/low level; red point: class 2/high level; black dashed  line: longitudinal trajectory which stayed in class 1; red  dashed  line: longitudinal trajectory which stayed in class 2; black full line: fitted trajectory which stayed in class 1; red  full line: fitted trajectory which stayed in class 2;  green dotted line: survival probability from proposed JLCM; blue dotted line: survival probability from basic JLCM; purple vertival dashed line: subject censored; red  vertical dotted line: time to event happened)}
		\end{center}
	\end{figure}

\end{document}